\documentclass[aps,prb,10pt,showpacs,twocolumn,amsmath,amssymb,superscriptaddress,floatfix]{revtex4-1}
\usepackage{graphicx}
\usepackage{amssymb, dsfont}
\usepackage{multirow}
\usepackage{subfigure}		
\usepackage{epstopdf}
\usepackage{graphicx}
\usepackage{epsfig}
\usepackage[breaklinks]{hyperref}

\def\beq{\begin{align}}
\def\eeq{\end{align}}
\newcommand{\bra}[1]{ {\langle{#1}|} }
\newcommand{\ket}[1]{ {|{#1}\rangle} }
\newcommand{\braket}[2]{ {\langle{#1}|{#2}\rangle} }
\newcommand{\norder}[1]{ {\mkern1mu\colon\mkern-4mu{#1}\colon\mkern-3mu} }
\newcommand{\xp}[1]{ {\left\langle{#1}\right\rangle} }
\newcommand{\bbone}{ {\mathds{1}} }
\newcommand{\gsd}{ {\mathfrak{m}} }

\usepackage[usenames,dvipsnames]{color}			
\definecolor{purple}{rgb}{0.5,0,0.5}
\newcommand{\roger}[1]{ { \color{purple} \footnotesize (\textsf{RM}) \textsf{\textsl{#1}} } }
\newcommand{\mike}[1]{ { \color{blue} \footnotesize (\textsf{MZ}) \textsf{\textsl{#1}} }}
\renewcommand{\roger}[1]{} \renewcommand{\mike}[1]{}

\begin{document}

\title{Exact matrix product states for quantum Hall wave functions}
\author{Michael P. Zaletel}
\affiliation{Department of Physics, University of California, Berkeley, California 94720, USA}
\author{Roger S. K. Mong}
\affiliation{Department of Physics, University of California, Berkeley, California 94720, USA}
\affiliation{Department of Physics, California Institute of Technology, Pasadena, California 91125, USA}

\begin{abstract}
We show that the model wave functions used to describe the fractional quantum Hall effect have exact representations as matrix product states (MPS).
These MPS can be implemented numerically in the orbital basis of both finite and infinite cylinders, which provides an efficient way of calculating arbitrary observables.
We extend this approach to the charged excitations and numerically compute their Berry phases.
Finally, we present an algorithm for numerically computing the real-space entanglement spectrum starting from an arbitrary orbital basis MPS, which allows us to study the scaling properties of the real-space entanglement spectra on infinite cylinders.
The real-space entanglement spectrum obeys a scaling form dictated by the edge conformal field theory, allowing us to accurately extract the two entanglement velocities of the Moore-Read state.
In contrast, the orbital space spectrum is observed to scale according to a complex set of power laws that rule out a similar collapse.
\end{abstract}


\maketitle

\section{Introduction}
The fractional quantum Hall (FQH) effects are exotic phases of matter that appear when interacting 2D systems are subject to large magnetic fields.
They are the foremost example of \emph{topologically ordered} phases, which are characterized by long range entanglement rather than by local order parameters.\cite{XGWenReview95}
Topological order has many signatures such as gapless edge excitations, fractional or non-abelian statistics, and ground state degeneracy on a cylinder or torus.
Many of these properties were discovered or demonstrated using ``model wave functions'' as ansatz for the ground state.
The first example was Laughlin's wave function\cite{Laughlin83} at filling $\nu = 1/q$ argued to explain the first FQH experiments,\cite{Tsui-1982} which has since been followed by many other successful ansatz.\cite{Halperin:FQHHierarchy:1983, Haldane:FQHHierarchy:1983, MooreRead91, ReadRezayi99}
The model wave functions have also served as a diagnostic for exact diagonalization (ED) studies by checking the model states' overlap with the ED ground state. 

Recently new ideas originating from quantum information, such as the entanglement spectrum, have become important tools for detecting and characterizing the topological order of these phases.\cite{KitaevPreskill:TEE06,LevinWen2006,AguadoVidal2008}
Given a bipartition of the system into two sub-Hilbert spaces, $\mathfrak{H} = \mathfrak{H}_A \otimes \mathfrak{H}_B$, we can decompose any wave function $\ket{\Psi}$ in terms of wave functions which live solely in $A$ or $B$:
\begin{align}
	\ket{\Psi} &= \sum_a e^{-\frac{1}{2}E_a}\ket{\Psi^A_a} \otimes \ket{\Psi^B_a}	,
	\label{eq:entspec}
\end{align}
with the restriction that the `entanglement spectrum' $E_a$ is real and that the `Schmidt vectors' $\ket{\Psi^A_a}$ form an orthonormal set (as do the $\ket{\Psi^B_a}$).
It was suggested in Ref.~\onlinecite{KitaevPreskill:TEE06}, and later thoroughly investigated in Ref.~\onlinecite{HaldaneLi2008} that when $A, B$ are chosen to be regions in space, the low-lying entanglement spectrum of a FQH state can be identified with the energy spectrum of the conformal field theory (CFT) describing its gapless edge excitations.\cite{QiKatsuraLudwig2012, Papic2011}
It was observed that for certain model wave functions, such as the Moore-Read (MR) state,\cite{MooreRead91} the \emph{entire} entanglement spectrum could be identified as states of the edge CFT.\cite{HaldaneLi2008}

	A second realm in which entanglement has come to play an important role is for a set of variational wave functions called `matrix products states' (MPS) \cite{FannesNachtergaeleWerner1992} in one-dimension (1D) or `tensor networks' \cite{VerstraeteCirac2004} in higher dimensions.
These are the variational states of the highly successful density matrix renormalization group (DMRG) method,\cite{White1992, OstlundRommer1995} which succeeds because MPS efficiently capture the structure of entanglement in many body wave functions.\cite{Perez-Garcia:2007}
The precise relationship between topological order and tensor network representations is a subject of ongoing work, but in 1D at least a complete classification of symmetry protected topological (SPT) order for both gapped 1D spin and fermion chains was recently accomplished using the MPS representation of the ground state. \cite{ChenGuWen2011, FidkowskiKitaev2011, TurnerPollmann, Schuch2011}
Given a set of sites labeled by $i$, each with local basis $\ket{m_i}$, an MPS $\ket{\psi}$ is defined by a set of `$B$-matrices',
\begin{align}
	\label{eq:defMPS}
	\ket{\psi} &= \sum_{\{ \alpha, m \}} \left[ \cdots B^{m_2}_{\alpha_{3} \alpha_2} B^{m_1}_{\alpha_{2} \alpha_1}  \cdots \right]\ket{ \cdots, m_2, m_1, \cdots }.
\end{align}
The indices $0 \leq \alpha_i < \chi$ to be traced over are called `auxiliary' indices, which we consider to be states in an `auxiliary Hilbert space' defined on the bonds between sites.
With the proper normalization, the auxiliary states are in one to one correspondence with the entanglement spectrum of a cut on the bond.
An important insight from the classification scheme is that a suitable renormalization procedure \cite{VCLRW2005, ChenGuWen2011} can be defined which produces a representative state of the smallest possible $\chi$.
For example, the $\chi = 2$ state of Affleck, Lieb, Kennedy and Tasaki (AKLT) \cite{AKLT} is representative of the SPT ordered Haldane phase \cite{Haldane1983} of the spin-1 Heisenberg chain.

	The observed simplicity of the FQH model states' entanglement spectrum suggests they play an analogous role for the FQH effects as the AKLT state does for the Haldane phase.
To pursue the analogy further, the 1+1D AKLT wave function can be written as a time ordered correlation function of a single `$0 + 1$D' spin-$\frac{1}{2}$, which leads to its simple expression as an MPS whose $\chi = 2$ auxiliary Hilbert space is a spin-$\frac{1}{2}$.\cite{ArovasAuerbachHaldane1988}
The 2+1D FQH model wave functions can be written as the correlation function of a 1+1D CFT.
Does it follow that the model FQH states have exact representations as an MPS with an auxiliary Hilbert space in one to one correspondence with the CFT, and if so,  can they be implemented and manipulated numerically?

In this paper we show that the model FQH wave functions and their quasiparticle excitations indeed have exact representations as MPSs.
As expected the requisite structure of the model states is that their wave functions are the correlation functions of a 1+1D CFT, which implies essentially by definition that they are MPSs whose auxiliary Hilbert space is the CFT.
We also explain how the edge excitations and ground state degeneracy arise in the MPS picture.

Working on a cylinder in the Landau gauge, we can view the system as a 1D chain of orbitals for which the $B$-matrices of Eq.~\eqref{eq:defMPS} are the matrix elements of  local operators of the CFT.
We have implemented these MPSs numerically for both the fermionic Laughlin and Moore-Read states on the geometry of an \emph{infinitely} long cylinder of circumference $L$, allowing us to measure arbitrary real-space correlation functions using the standard infinite MPS algorithms.
The infinite cylinder has a number numerical advantages, including the absence of boundaries, full translation invariance and no  curvature effects.
Compared to the torus geometry,\cite{LauchliBergholtz2010, LauchliBergholtzHaque2010, ZhaoBergholtz2012}
	only a single cut is required to study the entanglement, greatly simplifying the identification of the entanglement spectrum.
As we show in Sec.~\ref{sec:convergence}, the computational complexity of the MPS representation is on the order $\mathcal{O}(b^L)$ for $b \sim \mathcal{O}(1)$.
However, to achieve the same type of scaling in the traditional Hilbert space representation (say on a sphere\cite{Haque2007, ZozulyaFQHEntanglement07, HaldaneLi2008, ThomaleSterdyniak2010, Sterdyniak2012, Dubail2012a}) would require $N \sim \mathcal{O}(L^2)$ particles and a Hilbert space dimension scaling as $b^{L^2}$.
We note that previously a conceptually distinct approach found an MPS for the Laughlin state in which there is one matrix per \emph{particle}, rather than per orbital.\cite{Iblisdir2007}
However, the construction does not easily generalize to other FQH states and again results in a complexity $b^N$, which implies it cannot be implemented on the infinite cylinder geometry. 
A 2D tensor network construction for observables has also been constructed for lattice FQH states,\cite{BeriCooper2011} but the feasibility of its implementation is unclear.


Furthermore, we introduce an algorithm for calculating the real-space entanglement spectrum of any state given as an MPS in the orbital basis.
We first use the larger system sizes provided by the MPS representation to extract the topological entanglement entropy (TEE) $\gamma$ using four different methods; first using the conventional scaling of the entanglement entropy,\cite{KitaevPreskill:TEE06,LevinWen2006}
\begin{equation}
	S = \sum_a E_a e^{-E_a} = a_S^{\,} L - \gamma + \mathcal{O}(L^{-1}),	
\end{equation}
and second from a similar scaling form we derive for the lowest entanglement energy, $E_0 = a_E^{\,} L - \gamma + \mathcal{O}(L^{-1})$, for both the orbital and real-space cuts.
	\footnote{The lowest entanglement energy $E_0$ is equal to the R{\'e}nyi entropy (Ref.~\onlinecite{RenyiEntropy}) $S_\infty$ at infinite order.}
We find using $E_0$ in the orbital cut converges most quickly, and as this form is applicable to other topologically ordered phases, it may prove useful in cases where small system sizes are a constraint.
We are able to definitively determine $\gamma$ using all four methods for the $\nu = \frac13, \frac15$ and $\frac17$ Laughlin states,
	as well as the $\nu = \frac12$ Moore-Read state (\textit{cf}.\ Tab.~\ref{tab:vel_gamma}), which proved difficult in previous studies.%
	\cite{Haque2007, ZozulyaFQHEntanglement07, LauchliBergholtzHaque2010, Sterdyniak2012}

Finally, we perform a detailed scaling analysis of the spectrum for both the orbital and real-space cuts.
During the final preparation of this work, a recent work \cite{Dubail2012b} has conclusively demonstrated earlier arguments that the real-space entanglement spectrum of the model wave functions takes the form of the chiral Hamiltonian $H$ perturbed by local, irrelevant boundary operators.\cite{QiKatsuraLudwig2012, Dubail2012a} 
This implies a scaling collapse of the entanglement spectrum in the limit $L \to \infty$ for fixed CFT level $n$.
The large system sizes available here give the first detailed demonstration of this principle, allowing us to extract the entanglement velocities for both the Majorana and $\mathrm{U(1)}$ modes of the MR state, as well as the form of the leading irrelevant corrections.

In contrast, in the orbital cut each entanglement eigenvalue scales as $E_a - E_0 \sim L^{-\zeta_a}$ which precludes the possibility of collapsing the spectrum.
This is contrary to earlier indications that the orbital spectrum showed the same linear dispersion, \cite{ThomaleSterdyniak2010} though the case  studied there was the `conformal limit' of bosonic $\nu = \frac{1}{2}$ wave function on a finite sphere.

\begin{figure}[t]
	\includegraphics[width=0.35\textwidth]{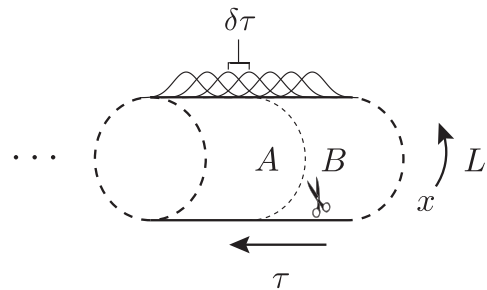}
	\caption{%
		The infinite cylinder geometry for model FQH wave functions.
		$L$ is the circumference, $x$ and $\tau$ are the coordinates around and along the cylinder respectively.
		$\delta \tau = \frac{2 \pi \ell_B^2}{L}$ is the spacing between Landau level orbitals, where $\ell_B = (\hbar/qB)^{1/2}$ is the magnetic length.
		A real-space entanglement cut between regions $A, B$ would be made along some fixed $\tau$. 
	}
	\label{fig:geometry}
\end{figure}

\begin{table}[t]
	\begin{align*} \begin{array}{l @{} c c c @{} c @{} }
		&	\textrm{Filling}	&	v_\phi	&	v_\chi & \mathcal{D}^2 = e^{2\gamma}	\\
	\hline\hline
		\multirow{4}{*}{\textrm{Laughlin}}
		&	1	&	2.2568 \pm 0.0003 & &	|\gamma| < 10^{-7}	\\
		&	1 / 3	&	1.2956 \pm 0.0006	& &	2.996	\\
		&	1 / 5	&	0.672 \pm 0.009	& &	4.96	\\
		&	1 / 7	&	0.28 \pm 0.02	& &	6.88	\\
	\hline
		\textrm{Moore-Read}
		&	1 / 2	&	1.33 \pm 0.01	&	0.21	\pm 0.01	 &	7.77 \\
	\hline\hline
	\end{array} \end{align*}
	\caption{%
		Extracted real-space entanglement velocities and TEE $\gamma$ for various model wave functions.
		$v_\phi$ is the velocity of the chiral boson, and for the MR case, $v_\chi$ the velocity of the chiral Majorana, in units of the magnetic length $\ell_B$.
		For the $\nu = 1$ integer quantum Hall state, the exact value of the velocity is known to be $4 / \sqrt{\pi}$.
		In the column for total quantum dimension $\mathcal{D}$, we present the value extracted via the orbital cut $E_0$ around $L = 25\ell_B$.
		(For $\nu = 1$ integer case, we use the real-space cut $S$ instead.)
		Refer to Sec.~\ref{sec:orb_vs_rs} for details on our numerical methods.
	}
	\label{tab:vel_gamma}
\end{table}

\section{Model Wavefunctions and Matrix Product States}

	A number of gapped model wave functions, including those of the FQH, can be written as the correlation functions of a field theory in one lower dimension.\cite{ShankarVishwanath2011}
In the 2+1D FQH effect, for example, the model wave functions are correlation functions of a 1+1D chiral conformal field theory (CFT).\cite{MooreRead91}
Other examples with this structure include the AKLT states, the Toric code,\cite{KitaevQC03} spin chains,\cite{CiracSierra-iMPS2010,Nielsen-BosonVertexOpMPS2011} and certain BCS superconductors.\cite{Volovik:HeDroplet,ReadGreen:p+ipFQHE00,DubailRead2011,ShankarVishwanath2011}
As we will illustrate in the case of the FQH effect, this structure implies that the state has an exact implementation as an MPS or a tensor network.
The auxiliary Hilbert space of the tensor network is in correspondence with the Hilbert space of the associated lower dimensional field theory.
In turn, the edge excitations and the entanglement spectrum of tensor networks are known to be closely related; \cite{PhysRevB.83.245134} this relationship takes a particularly elegant form for the FQH effect due to the stringent constraints of conformal invariance in 2D.\cite{Dubail2012b}

	The simplest example is the Laughlin state on an infinite cylinder, which can be written as the correlation function of a chiral boson $\phi(z)$ 
	\footnote{The normal ordering prescription is as follows. Each insertion $\norder{e^{i \sqrt{q} \phi(z_a)}}$ is normal ordered, which eliminates a contraction that would produce $(z_a - z_a)^q = 0$, and the self interaction of the background charge $\int\!d^2z\,\phi(z)$ is ignored, which would contribute an overall divergent constant.}
\begin{align}
\label{firstquant}
	\Psi_L(z_a) &= \prod^N_{a < b} \sin\big( (z_a - z_b) \tfrac{\pi}{L}\big)^q e^{ -\frac{1}{2 \ell_B^2} \sum_a\!\tau_a^2 } \\
		& = \left< \exp\bigg[ {i \sqrt{q} \sum_{a=1}^N \phi(z_a) - i \sqrt{q} \rho \int\!d^2z\,\phi(z)} \bigg] \right>_\phi	\notag
\end{align}
as elucidated by Moore and Read.\cite{MooreRead91}
Throughout we will use $z = x+i\tau$ as a complex coordinate on the cylinder, where $x$ runs around its circumference of length $L$ and $\tau$ runs along its length, as illustrated in Fig.~\ref{fig:geometry}.
The filling fraction is $\nu = \frac{p}{q}$, the magnetic length is $\ell_B^2 = \frac{\hbar}{e B}$, and the density of electrons is $\rho = \frac{\nu}{2 \pi \ell_B^2}$.
	The chiral boson $\phi$ is a free field characterized by its correlation function on the plane or cylinder,
\begin{align}
	\xp{ \phi(z) \phi(z') }_\textrm{plane} &= - \log( z - z')	,	\notag \\
	\xp{ \phi(z) \phi(z') }_\textrm{cyl} &= - \log\, \sin\left[\tfrac{\pi}{L} (z - z') \right]
		.
\end{align}

In the Laughlin state, for each electron we insert the operator $\mathcal{V}(z_a) = \norder{ e^{i \sqrt{q} \phi(z_a)} }$, where $\norder{}$ denotes normal ordering.
Other quantum Hall states, such as the Moore-Read state or the Read-Rezayi sequence,\cite{MooreRead91, ReadRezayi99} can be obtained by letting $\mathcal{V}$ be an operator in a more general CFT.
It is also necessary to include a neutralizing `background charge' $\mathcal{O}_{bc} = - i  \rho \int\!d^2z \,\phi(z) / \sqrt{\nu} $.
The background charge introduces some subtleties, as the branch cut in the bosonic propagator has a phase ambiguity equivalent to a choice of gauge for the electrons, which we will address at a later point.


	We write a second quantized version of Eq.~\eqref{firstquant} using a coherent state wave function in the variable $\psi$ (which is a complex/Grassmann number for bosons/fermions), which in the thermodynamic limit is
\begin{align}
	\Psi_L[\psi] &= \bra{0} e^{\int\!d^2z \, \psi(z) \hat{\Psi}(z) } \ket{\Psi_L}		\notag\\
	 &= \left< e^{
		\int\!d^2z \left[\mathcal{V}(z) \, \psi(z) -  i  \rho \phi(z) / \sqrt{\nu} \right]
	} \right>_\textrm{CFT}
	. \label{eq:secondquant}
\end{align}
The notation is rather subtle as we are tying together two theories: the physical particles in 2+1D, with coherent state coordinate $\psi(z)$, and the path integral over the auxiliary space of the 1+1D CFT, characterized by the correlation functions  $\left< \cdot \right>_\textrm{CFT}$.
Number conservation is enforced by the $\mathrm{U(1)}$ symmetry of the chiral boson.

	The structure of Eq.~\eqref{eq:secondquant} is identical to that of a `continuous matrix product state' (cMPS) defined in Ref.~\onlinecite{PhysRevLett.104.190405}, which we review briefly.
Starting with an MPS for a chain of bosons or fermions at sites with positions $\tau$, we first pass from the occupation basis $\{\ket{m_\tau}\}$ to the coherent state basis $\{\ket{\psi_\tau}\}$ by defining
\begin{equation}
	B_{\alpha \alpha'}[\psi_\tau] \equiv \sum_{m_\tau}  \bra{\psi_\tau}B^{m_\tau}_{\alpha \alpha'} \ket{m_\tau}.
\end{equation}
Second, we note that the trace over the auxiliary states $\{\alpha\}$ is formally equivalent to a path integral over a 1D system, with the $B$ playing the role of transfer matrices.
Anticipating the continuum limit, we assume there are matrices $H, V$ in the auxiliary Hilbert space such that $B[\psi(\tau)] = e^{H(\tau) + V(\tau) \psi(\tau)}$.
We can then take the continuum limit of the MPS by analogy to the usual time-ordered path integral, which defines a cMPS,
\begin{equation}
	\label{cMPS}
	\Psi[\psi] = \operatorname{Tr}_\textrm{aux} \left[ \mathcal{T} e^{\int_0^{L_\tau}\!d\tau \left[ H(\tau) + V(\tau) \psi(\tau) \right]} \right].
\end{equation}
Comparing the cMPS to the second quantized version of the Laughlin state, \eqref{eq:secondquant}, we see that they are equivalent if we take the \emph{physical} Hilbert space at each slice to be that of particles on a ring of circumference $L$, and the auxiliary Hilbert space to be that of a chiral boson.
In this case, $H$ is precisely the Hamiltonian of the CFT (plus the background charge $\mathcal{O}_{bc}$), while $V$ is the electron operator, $\mathcal{V}$.
This structure was also recently noted in Ref.~\onlinecite{Dubail2012b}, where, in the MPS language, they find the dominant eigenvector of the `transfer matrix'\cite{Perez-Garcia:2007} of the cMPS, from which the entanglement Hamiltonian follows.

	While the cMPS representation is convenient from an analytic perspective, computationally it is desirable to have the discrete version expressed in the basis of lowest Landau level (LLL) orbitals.
Defining a coordinate $w = e^{-\frac{2 \pi i}{L}z}$ for notational convenience, in Landau gauge the orbitals can be written as
\begin{align}
	\label{eq:LLL}
	\varphi_n(z) \propto  e^{- \frac{1}{\ell_B^2 }\left(i \tau_n x + \frac{1}{2} (\tau - \tau_n)^2 \right)}
		= w^n e^{-\frac{1}{2 \ell_B^2} \left( \tau_n^2 + \tau^2\right) }	,
\end{align}
where $\tau_n=\frac{2 \pi n}{L} \ell_B^2$ is the guiding center for the $n$\textsuperscript{th} orbital.
Viewing the orbitals as the sites of an infinite 1D chain, we want to arrive at a discrete MPS as defined in \eqref{eq:defMPS}, which requires finding the appropriate matrices $B^m_{\alpha \alpha'}$.
Based on the cMPS, we expect $\alpha$ will be in one-to-one correspondence with the states of the associated CFT.

	In order to extract the occupation number at orbital $n$, we take advantage of the fact that orbitals of the LLL (in the Landau gauge) are labeled by momentum.
When acting on a many-body state in the LLL, we can then replace the destruction operator $\hat{\psi}_n$ for orbital $n$ with a contour integral around the cylinder,
\begin{equation}
	\hat{\psi}_n \longrightarrow  e^{ \frac{\tau_n^2}{\ell_B^2} }  \!\!\underset{\tau = \tau_n}{\oint}\!
		\frac{dw}{2 \pi i}\,  w^{-n-1} \hat{\psi}(w)	.
\end{equation}
We chose to perform the integrals at $\tau_n$, though with an appropriate change in normalization a different location could be chosen.
	
\begin{figure}[t]
	\subfigure[Structure of the orbital MPS.]{ \label{fig:mps_chain} \includegraphics[width=0.40\textwidth]{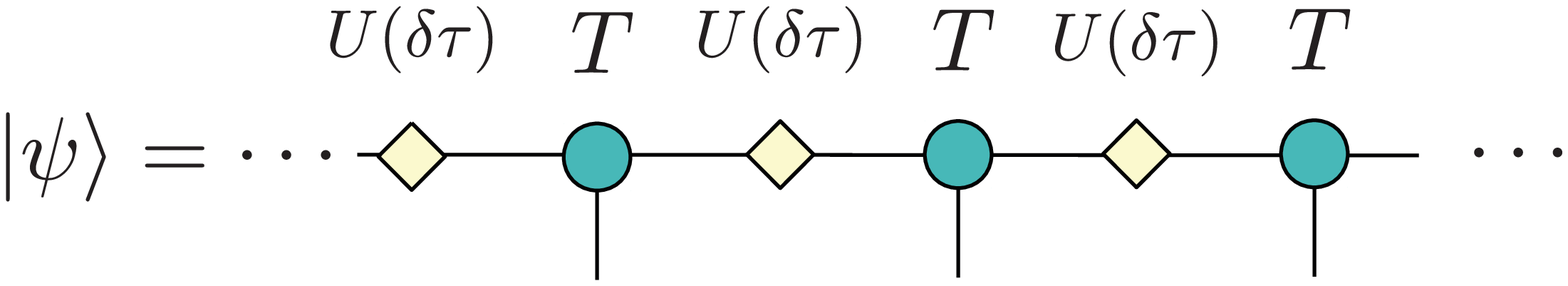} }
	\subfigure[Definition of the $B$-matrix.]{ \label{fig:mps_B} \quad\includegraphics[width=0.22\textwidth]{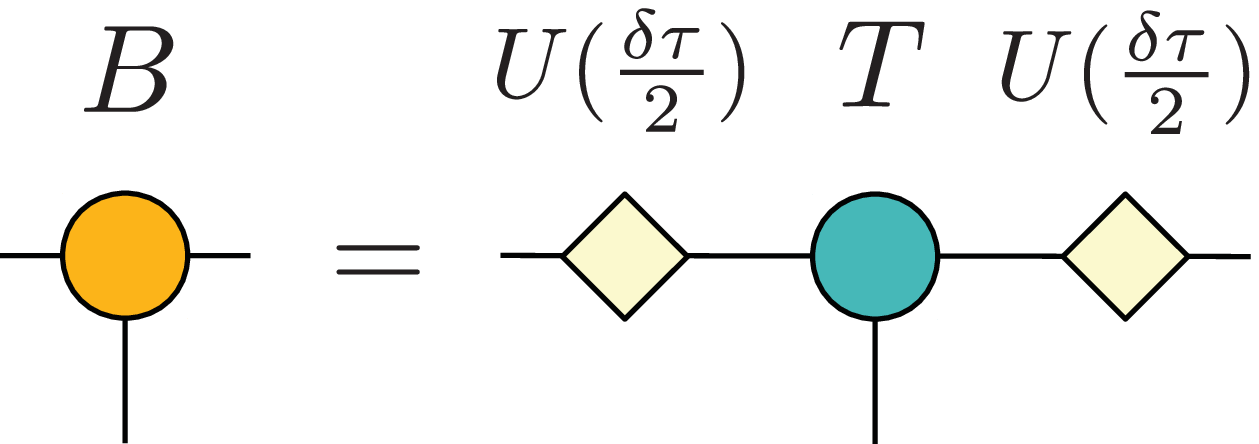}\quad }
	\subfigure[Structure of MPS with a quasihole insertion $Q$.]{ \label{fig:mps_withH} \includegraphics[width=0.40\textwidth]{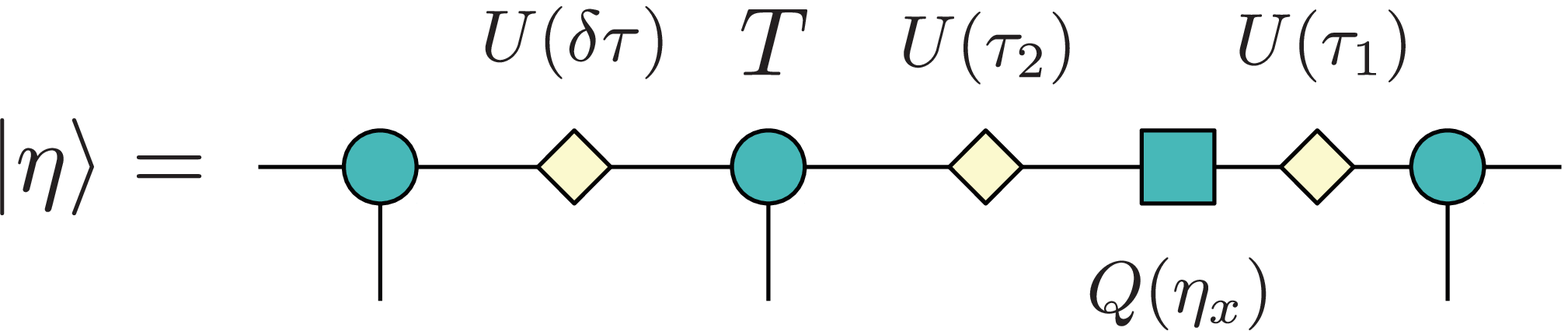} }
	\caption{The structure of the orbital MPS.
		(a) $U$ is free time evolution of the CFT, punctuated by perturbations $T$ at $\tau = \tau_n$.
		(b) The $B$-matrices in \eqref{eq:defMPS} are defined by combining $U$ and $T$.
		(c) A quasiparticle is inserted by placing the matrix elements of the vertex operator $Q$ in the correct time-ordered positions.
			It can be absorbed into either of the adjoining $B$-matrices.
	}
	\label{fig:mps}
\end{figure}

	The gauge of the cMPS, however, depends on a branch cut prescription for the background charge.
It is convenient to choose the cut to consistently occur at some fixed $x$ coordinate, such as the boundary of $-L/2 < x < L/2$.
This choice of gauge does not produce the Landau gauge; they differ by a phase $e^{i x \tau  \ell_B^{-2}}$. Choosing this branch cut prescription for $\phi(z)$, but keeping $\psi_n$ to be the destruction operators for the Landau gauge, we find Eq.~\eqref{eq:secondquant} can be brought to the form
\begin{align}
	\label{eq:orbitalMPS}
	\Psi[\psi_n] &= \left\langle
		e^{ \sum_n \!\!\underset{\tau = \tau_n}{\oint}\hspace{-1ex} \frac{dw}{2 \pi i} \, w^{-1} \left[ \mathcal{V}(w) \psi_n  -  i \sqrt{\nu} \phi(w) \right] }
		\right\rangle_\textrm{CFT}.
\end{align}
Eq.~\eqref{eq:orbitalMPS} looks like unperturbed time evolution governed by the Hamiltonian of the chiral CFT, $H$, punctuated by interactions at $\tau_n$.
As orbital ordering coincides with time ordering, we can pass to the Hamiltonian picture by inserting resolutions of the identity $\openone = \sum_\alpha \ket{\alpha}\bra{\alpha}$ at positions $\tau = \tau_n \pm \epsilon$, where $\alpha$ label all states of the CFT.
A resulting unit cell looks like
\begin{align}
	\label{eq:orbitalchain}
	\Psi[\psi_n]  =  \sum_{ \{\alpha\} }&
		\bigg[ \cdots \, \bra{\alpha_{n+1}} e^{-\delta \tau H} \ket{\alpha_{n}}   \notag\\
		 &\quad  \bra{\alpha_{n}} e^{ \mathcal{V}_0 \psi_n - i \sqrt{\nu} \phi_0 } \ket{\alpha_{n-1}} \, \cdots \bigg]	,
\end{align}
where the operator
\footnote{While we have expressed the definition in terms of the coordinate $w$, we are taking only a Fourier mode on the cylinder, not a conformal transformation to the plane. If we were to conformally map the expression to the plane we would recover the familiar modes of radial quantization, but that is not necessary here.}
\begin{equation}
	\hat{\mathcal{V}}_{0} \equiv \oint \frac{dw}{2 \pi i} \,  w^{-1}  \mathcal{V}(w) \notag
\end{equation}
is precisely the `$0$\textsuperscript{th} mode' of the electron operator $\mathcal{V}(w)$,
and likewise $\phi_0$ is the zero-mode of the chiral boson.
The resulting transfer operators are of two types.
For the unperturbed segments $\tau \in (\tau_n, \tau_{n-1})$, the transfer operator is
\begin{equation}
	U(\delta \tau) \equiv e^{- \delta \tau H}, \quad U(\delta \tau)_{\alpha \beta} = \delta_{\alpha \beta}e^{- \delta \tau E_\alpha }
	\label{eq:defU}
\end{equation}
where $\alpha$ again runs over states of the CFT, with energies $E_\alpha$, and $\delta \tau = \frac{2 \pi}{L} \ell_B^2$.
At the location of each site we define a transfer operator
\footnote{Note that the boundary condition of the CFT on the auxiliary bond is not necessarily periodic, due to the zero mode which `twists' the boundary condition. }
\begin{align}
	T_{\alpha \beta}[\psi_n] &\equiv \bra{\alpha}e^{ \, \hat{\mathcal{V}}_{0} \psi_n  -   i \sqrt{\nu} \phi_0} \ket{\beta}.
\end{align}
Stringing the transfer matrices together, we arrive at the exact MPS,
\begin{equation}
	\Psi[\psi_n] = \prod_n U(\delta \tau) T[\psi_n]
\end{equation}
as illustrated in Fig.~\ref{fig:mps_chain}.
We have suppressed the implicit summation over the CFT states $\alpha$.

The above is in `coherent state' form; to convert to the occupation basis $\{\ket{m}\}$, we define the $B$-matrices of Eq.~\eqref{eq:defMPS} to be
\begin{align}
	\sum_m  B^m (m!)^\frac{3}{2}  \psi^m
		&\equiv U(\tfrac{1}{2}\delta \tau) \, T[\psi] \, U(\tfrac{1}{2}\delta \tau)
\end{align}
as shown in Fig.~\ref{fig:mps_B}.
Explicitly,
\begin{align}
\label{eq:Bm}
	B^m &= 	 U(\tfrac{1}{2}\delta \tau) e^{-  i \sqrt{\nu} \phi_0/2}  \frac{\big(\hat{\mathcal{V}}_0\big)^m}{\sqrt{m!}} e^{- i \sqrt{\nu}\phi_0/2} U(\tfrac{1}{2}\delta \tau).
\end{align}
While the result is general, for the Laughlin and Moore-Read states, which  are described by free CFTs with electron operators
\begin{subequations}\begin{align}
	\label{vertexOps}
	\mathcal{V}(w) &= \norder{ e^{i \sqrt{q} \phi(w)} }  \, \, \textrm{(Laughlin)}, \\
	\mathcal{V}(w) &= \chi(w) \norder{ e^{i \sqrt{q} \phi(w)} } \, \, \textrm{(Moore-Read)},
\end{align}\end{subequations}
($\chi$ is a chiral Majorana field),  both $T$ and $U$, and hence $B$, can be calculated exactly at negligible numerical cost (for the details of this calculation, we refer to Appendixes~\ref{app:eval_B} and~\ref{app:eval_B_MR}).
For an arbitrary CFT, their calculation is more involved but nevertheless tractable using  formulas developed for 
the `truncated conformal space' approach to perturbed CFTs.\cite{YurovZamolodchikov}

	In summary, we have demonstrated how to take a model wave function written in terms of a correlator of a CFT and convert it to a discrete MPS [Eq.\eqref{eq:defMPS}] in the orbital basis, characterized by a set of $B^m_{\alpha \beta}$.
The auxiliary indices $\alpha, \beta$ label states of the CFT, such that each matrix $B^m$ is an operator of the CFT.
The operator $B^m$ consists of three pieces: the (imaginary) time-evolution of the CFT ($U$), the background charge ($e^{-i \sqrt{\nu} \phi_0}$), and the insertion of $m$ electron operators ($\hat{\mathcal{V}}_0^m$). This is the chief result of this paper.

\subsection{Discussion}
	In order to obtain wave functions on a half or finite cylinder, one simply truncates the MPS using the vacuum of the CFT as a boundary condition for the severed auxiliary bonds.
If excited states are used as the boundary condition, these produce the corresponding model edge excitations.\cite{Wen1992}
This structure is analogous to the spin-$\frac{1}{2}$ degree of freedom at the boundary of an AKLT chain, which arises from the two choices of boundary condition for the $\chi = 2$ MPS.

	To understand the ground state degeneracy of the phase, note that if the phase is $\gsd$-fold degenerate on an infinite cylinder, there are $\gsd$-primary fields in the CFT, and the states of the CFT partition into $\gsd$ families which `descend' from each of these primary fields.\cite{FendleyFisherNayak2007}
Each family is invariant under the action of the electron operator $\mathcal{V}(w)$, so it follows that the CFT states on a given auxiliary bond can be consistently truncated to one of these $\gsd$ families.
The $\gsd$ choices on the bond provide the $\gsd$ `minimal entanglement states'.\cite{TBONE}
	
	The `thin torus' wave functions are also a limiting case of our construction.\cite{TaoThouless1983, Bergholtz2006}
As $L \to 0$, we can truncate the MPS by keeping only the states of the CFT with the lowest energy within each family (the `highest weight states'), which generates a $\chi = 1$ MPS. 
The construction intuitively connects how the operator product expansions in the CFT are related to the orbital occupation numbers in the thin torus limit, the so-called ``pattern of zeros''\cite{WenWang-2008} or the ``root configuration.''\cite{BernevigHaldane2008}
We also note that an approximate $\chi=2$ MPS for the Laughlin state was recently found;\cite{NakamuraWangBergholtz2012} in our language this results from a truncation of the CFT to the states $|P|\leq1$.

	We now explain how the two conserved quantities of the LLL problem, particle number and momentum (sometimes called `center of mass'), can be assigned to the states of the CFT.
In the orbital basis we define the conserved quantities to be
\begin{subequations}\begin{align}
	\hat{C} &= \sum_{j} ( q \hat{N}_j - p ) \quad \textrm{(particle number)}	, \\
	\hat{K} &= \sum_{j} j ( q \hat{N}_j - p ) \quad \textrm{(momentum)}	,
\end{align}\end{subequations}
where $j$ is the orbital index and we have included a filling factor dependent scaling  ($\nu = p/q$) so that both remain finite in the thermodynamic limit.
If a state is invariant under a $\mathrm{U(1)}$ symmetry transformation, the states of the Schmidt spectrum can be assigned definite charge. Consequently, the entanglement spectrum on bond $\bar{n} \in \mathbb{Z} + \frac{1}{2}$ can be labeled by pairs $(C_{\bar{n}}, K_{\bar{n}})$.
The states of the auxiliary CFT have quantum numbers as well, in particular the total momentum $|P|$ of the CFT and the winding number $N$ of the boson (see Appendix \ref{app:eval_B} for detailed definitions).
The pairs $(N, |P|)$ and $(C_{\bar{n}}, K_{\bar{n}})$ are related by
\begin{subequations}\begin{align}
	C_{\bar{n}} &= N	, \\
	K_{\bar{n}} &= q |P| + \frac{1}{2} N^2 + \bar{n} N	,
\end{align}\end{subequations}
which explains how the previously observed offsets of the $|P| = 0$ levels depend on the number sector and bond location.

\section{Convergence properties and computational complexity}
\label{sec:convergence}
\begin{figure}[t]
	\includegraphics[width=77mm]{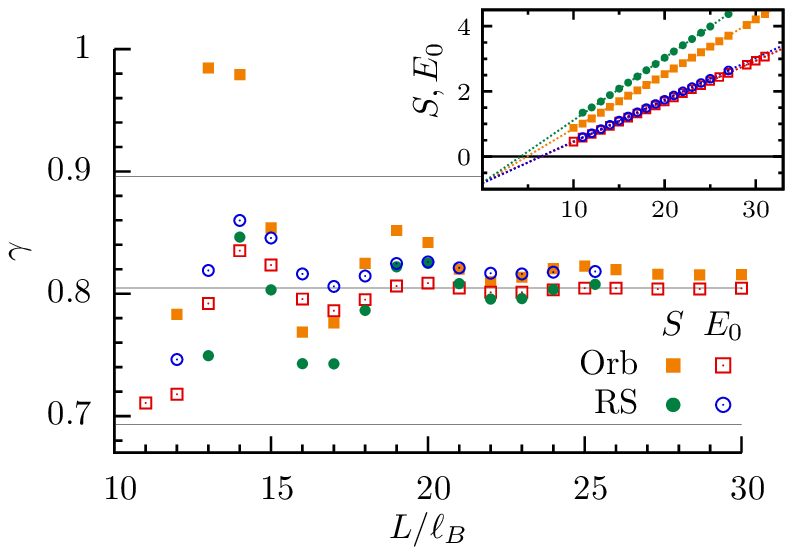}
	\\[3mm]
	\includegraphics[width=77mm]{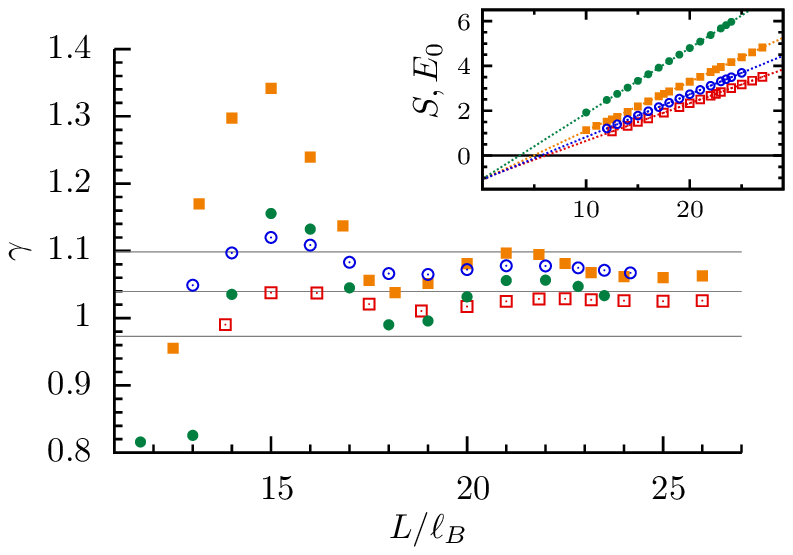}
	\caption{%
		Numerically computing the TEE $\gamma$ for the $\nu=1/5$ Laughlin state (top) and the $\nu=1/2$ Moore-Read state (bottom).
		$\gamma$ is extracted from both orbital (squares) and real-space (circles) cut, via the entanglement entropy $S$ (filled) and the lowest entanglement energy $E_0$ (empty),
			by performing windowed fits to the form $S(L),E_0(L) = a_{S,E}^{\,} L - \gamma$ at various circumferences $L$.
		The horizontal lines mark the values of $\gamma = \log \mathcal{D}$ where $\mathcal{D}^2 = 4,5,6$ (top) and $7,8,9$ (bottom).
		As $L \rightarrow \infty$, the extracted value of $\gamma$ approaches their theoretical values of $\frac{1}{2}\log5$ and $\frac{1}{2}\log8$ respectively.
		In the latter case we can see that $L \gtrsim 20\ell_B$ is required for the TEE to be extracted with reasonable accuracy.
		(Insets) $S$ vs.\ $L/\ell_B$ for the four cases.
	}
	\label{fig:gamma_fit}
\end{figure}
	For numerical purposes we must truncate the MPS by keeping only the $\chi$ most important states in the entanglement spectrum.
Most MPS algorithms (such as measuring correlation functions) can then be computed with time $\mathcal{O}(M \chi^3)$ and storage $\mathcal{O}(\chi^2)$, where $M$ is the number of sites involved in the measurement. 
In this section we argue that to simulate the state at some fixed precision we must keep $\chi \sim e^{\alpha c L/v} (c L/v)^{-1/2}$ where $c$ is the central charge of the entanglement spectrum, $v$ is its `entanglement velocity,' and $\alpha$ is a non-universal constant of order 1.
In contrast to exact diagonalization, the complexity scales exponentially only in the circumference of the cylinder, rather than its area.
Use of the conserved quantum numbers drastically reduces the computational time, but does not alter the exponential complexity.
	
	Following Kitaev and Preskill's derivation of the topological entanglement entropy,\cite{KitaevPreskill:TEE06} we proceed under the assumption that the `thermodynamic' properties of the entanglement spectrum, such as the entanglement entropy, take the same form as those of the auxiliary CFT.
However, there is no reason to expect the velocities that appear will be universal, so in what follows all powers of $L/v$ should be understood to have non-universal coefficients.
The exact status of this assumption for the orbital basis is somewhat unclear, because as we will show the orbital spectrum does not collapse to the CFT; nevertheless the appearance of $\gamma$ in the entropy $S$, the scaling form of the lowest eigenvalue $E_0$, and the collapse we find for the convergence of $S$ with increased $\chi$ appear to behave as expected. 
		
	The density of states $\rho(E)$ for a modularly invariant CFT is given by the `Cardy' formula.\cite{Cardy1986}
However, when working with the `minimally entangled'\cite{TBONE} ground states naturally provided by the MPS construction, we must take into account the fact that only one sector of the CFT, `$a$,' belongs to the entanglement spectrum, where the sector $a$ depends on a choice of one of the $\gsd$ ground states.
The corresponding partition function and density operator are defined as	
\begin{align}
	\mathcal{Z}_{a} &= \operatorname{Tr}_{a}  e^{-\beta H_e}	,\\
	\hat{\rho} &= \mathcal{Z}_{a}^{-1} \operatorname{Tr}_{a} e^{-H_e}	.
\end{align}
The derivation of the Cardy formula requires a modular transformation, but the required partition function is not modularly invariant.
This results in the explicit appearance of the modular $\mathcal{S}$ matrix, $-\log(\mathcal{S}^{\bbone}_{a}) = \gamma_a$, where $\gamma_a$ is the topological entanglement entropy of the ground state $a$.
Taking this term into account, the density of states is
\begin{align}
	\rho(E) dE &=  \frac{dE}{4 E} \sqrt{ \frac{2}{\pi}} e^{-\gamma_a} e^{\sqrt{ \frac{\pi (c + \bar{c}) E L }{3 v}}}  \left(  \frac{\pi (c + \bar{c}) E L }{3 v} \right)^{1/4}.
\end{align}
All the other thermodynamic properties follow from $\rho(E)$.
It is convenient to introduce the dimensionless variable $\mu$, 
\begin{align}
	\mu &\equiv \left(  \frac{\pi (c + \bar{c}) E L }{3 v} \right)^{1/4}	.
\end{align}
We can calculate the partition function and entanglement entropy,
\begin{align}
	\mathcal{Z}_a(\beta) &= \int\! \rho(E) e^{- \beta E} dE \notag\\
		&= \sqrt{ \frac{2}{\pi} } e^{-\gamma_a} \int e^{\mu^2 - \frac{3 v}{\pi (c + \bar{c})} \frac{\beta}{L} \mu^4} d\mu \notag\\
		&= e^{\frac{\pi (c + \bar{c})}{12} \frac{L}{\beta v} - \gamma_a + \dots}	,\label{eq:Z}\\ 
	S &= \partial_{\beta^{-1}} (-\beta^{-1} \ln \mathcal{Z}_a) \big\vert_{\beta = 1}	\notag\\
		&= \frac{\pi (c + \bar{c})}{6} \frac{L}{v} - \gamma_a + \dots .
\end{align}
The partition function \eqref{eq:Z} is evaluated via steepest descent about the saddle point $\mu_\ast = \sqrt{\frac{\pi (c + \bar{c})}{6 v} \frac{L}{\beta}}$.
As the entanglement spectrum is $p_i = e^{-E_i} / \mathcal{Z}_a(1)$, a particular consequence of Eq.~\eqref{eq:Z} is that the lowest entanglement level is $p_0 = e^{-\left[ a_E L - \gamma_a \right] } $ for some non-universal $a_E$.
A similar result was recently obtained in Ref.~\onlinecite{Dubail2012b}.
As illustrated in Fig.~\ref{fig:gamma_fit}, for both the orbital and real-space cuts $\gamma$ can be extracted from the scaling of $p_0$ with equivalent or better accuracy as from $S$.
\begin{figure}[t]
	\includegraphics[width=0.38\textwidth]{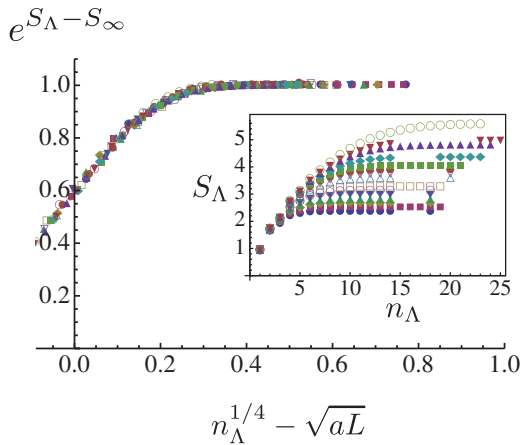}
	\caption{%
		Convergence of the orbital entanglement entropy $S_\Lambda$ for the $q = 3$ Laughlin state as the number of Virasoro levels kept ($n_\Lambda$) is increased.
		(Inset) For various circumferences $19\ell_B \leq L \leq 40\ell_B$, we calculate the entanglement $S_\Lambda$ of the MPS keeping only the lowest $n_\Lambda$ Virasoro levels of the CFT. For large enough $n_\Lambda$, $S_\Lambda$ converges to the exact entanglement entropy $S_\infty$.
		We expect the convergence to be controlled by the parameter $\mu - \mu_\ast \propto   n_\Lambda^{1/4} - (a L)^{1/2}$ for some $a$.
		(Main figure) We plot the convergence of the entanglement entropy, $e^{ S_\Lambda - S_\infty}$ as a function of $n_\Lambda^{1/4} - (a L)^{1/2}$, with $a \approx 0.0875$ giving a good collapse.
	}
	\label{fig:SvN} 
\end{figure}

	The steepest descent analysis shows that the bulk of the probability comes from a region within $\mathcal{O}(1)$ of the saddle point $\mu_\ast$.
Up to this point the number of states $\chi$ with $E < E_\ast$ is 
\begin{align}
	\chi(E_\ast) &= \int^{E_\ast}\!\!\!\rho(E) \, dE \,.
\end{align}
Alternately, we can define the number of CFT Virasoro levels $n_\ast$ required above the vacuum state,
\begin{align}
	n_\ast &\equiv n(E_\ast) = \frac{E_\ast L}{2 \pi v} = \frac{(c + \bar{c})}{24} \left(\frac{L}{v}\right)^2.
\end{align}
To study the convergence properties, suppose we only keep states such that $E < E_\Lambda$.
The cutoff partition function $\mathcal{Z}_\Lambda$ is
\begin{align}
	\mathcal{Z}_\Lambda(\beta) &= \int^{E_\Lambda}\!\! dE \, \rho(E) \, e^{- \beta E}		\notag\\
	&\sim \mathcal{Z}_\infty(\beta) \, \frac{1}{2} \operatorname{erfc} \big({-\sqrt{2}(\mu_\Lambda - \mu_\ast)}\big)	,
\end{align}
with resulting truncation error
\begin{align}
	\epsilon_\Lambda
	&\equiv 1 - \frac{\mathcal{Z}_\Lambda(1)}{\mathcal{Z}_\infty(1)}
	= \frac{1}{2} \operatorname{erfc}\big(\sqrt{2}(\mu_\Lambda - \mu_\ast)\big)	.
\end{align}
While the specific functional form may not remain universal, it suggests that convergence is controlled by the dimensionless factor $\mu_\Lambda - \mu_\ast \sim n_\Lambda^{1/4} - (a L)^{1/2}$.
In the inset of Fig.~\ref{fig:SvN}, we plot the convergence of the entanglement entropy $S_\Lambda$ as a function of the number of Virasoro levels $n_\Lambda$ kept at various circumferences $L$.
We then scale the data horizontally by plotting as a function of $n_\Lambda^{1/4} - (a L)^{1/2}$ for a numerically fit value of $a$.
Without any further \emph{vertical} scaling, the data appears to collapse.
This is somewhat surprising given the irregular structure of the orbital spectrum, but does validate the predicted form $n_\Lambda \sim L^2$.
Choosing an acceptable fractional error for $S_\Lambda$, in large $L$ limit we then conclude from the Cardy formula the required dimension of the MPS to simulate at fixed accuracy is
\begin{align}
	\chi  \sim e^{\alpha c L/v} (c L/v)^{-1/2}
\end{align}
as claimed.
Equivalently the number of Virasoro levels required is $\sim \mathcal{O}(L^2)$.

\section{Quasi-particle excitations}

	We now discuss how to introduce quasiparticles into the MPS.
In the `conformal block' approach to model wave functions,\cite{MooreRead91} a quasiparticle excitation at $\eta$ is introduced by inserting an appropriate operator $\mathcal{Q}(\eta)$ into the CFT correlator,
\begin{equation}
	\Psi[\psi_n; \eta] = \left\langle \mathcal{Q}(\eta)\,
		e ^ { \sum_n \left[ \mathcal{V}_0(\tau_n) \psi_n -  i  \sqrt{\nu} \phi_0(\tau_n) \right] }
		\right\rangle_\textrm{CFT}.
	\label{eq:qpmps}
\end{equation}	
We will focus on the Laughlin and Moore-Read quasiholes, for which $\mathcal{Q}$ is a \emph{local} operator that takes a particularly simple form,
\begin{subequations}\begin{align}
	\mathcal{Q}(\eta) &= \norder{ e^{i \phi(\eta)/\sqrt{q}} } \quad \textrm{(Laughlin)}, \\
	\mathcal{Q}(\eta) &= \sigma(\eta) \norder{ e^{i \phi(\eta)/2 \sqrt{q}} } \quad \textrm{(Moore-Read)}.
\end{align}\end{subequations}
Here $\sigma(\eta)$ is the chiral part of the Ising order operator.
Quasiparticles require `quasi-local' operators,\cite{PhysRevB.80.165330} which can also be included in the MPS, but we have deferred their implementation. 

To incorporate the quasihole into the MPS, we first explicitly time order Eq.~\eqref{eq:qpmps} by bringing the insertion $\mathcal{Q}(\eta)$ between the orbitals $\tau_{n+1} \geq \eta_\tau \geq \tau_n$.
For fermions, this introduces a sign for each electron in the region $\tau > \eta_\tau$.
As detailed in Appendix~\ref{app:eval_Q_qh}, this sign can be written as $s^{\hat{\pi}_0/\sqrt{q}}$, where $\hat{\pi}_0$ is conjugate to the bosonic zero-mode and $s = \pm 1$ for bosons and fermions respectively.

	We calculate the matrix elements of $\mathcal{Q}$ at $\tau = 0$,
\begin{align}
	Q_{\alpha, \beta} &= \bra{\alpha} s^{\hat{\pi}_0/\sqrt{q}}\mathcal{Q}(\eta_x)\ket{\beta}	,
\end{align}
and then insert $Q$ into the `unperturbed' evolution on the bond between sites $n, n+1$,
\begin{align}
	U(\delta \tau) &\to U(\tau_a) \, Q \, U(\tau_b), \, \,\, \tau_a + \tau_b = \delta \tau	,
\end{align}
where $\tau_a = \tau_{n+1} - \eta_\tau$.
The structure of the resulting MPS is illustrated in Fig.~\ref{fig:mps_withH}.
For further details on calculating $Q$ for the Laughlin and MR states we refer to Appendix~\ref{app:eval_Q_qh}.

\begin{figure}[t]
	\includegraphics[width=0.5\textwidth]{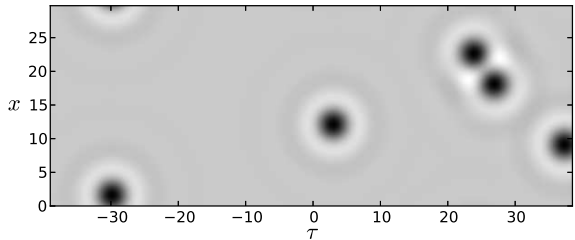} 
	\caption{%
		The real-space density $\rho(\tau, x)$ of a $q=5$ Laughlin state with five quasi-holes on an infinite cylinder of circumference $L = 30 \ell_B$. Distances are measured in units of $\ell_B$.
		(The top and bottom edges are identified.)
	}
	\label{fig:quasiholes}
\end{figure}

We have implemented the Laughlin quasiholes numerically, with a resulting density profile for a collection of quasiholes in the $q = 5$  state shown in Fig.~\ref{fig:quasiholes}. As a simple test of the result, we can explicitly evaluate the Berry connection associated with the transport of one $q = 3$ Laughlin quasihole around another,
\begin{align}
	\theta = \oint\! dA = \oint\! d\eta \, \bra{\eta}(-i\partial_\eta) \ket{\eta}.
\end{align}
We keep one quasihole fixed at $\eta = 0$, while a second follows a discretized path $\eta_i$ chosen to wind around the other, which defines a discretized connection $e^{i A_{ij}} = \braket{\eta_i}{\eta_j}$.
We then integrate the connection after subtracting out a similar phase in the absence of the second particle.
Calculating the inner product between two matrix product states can be computed with complexity $\mathcal{O}(A/\ell_B^2 \, \, \chi^3)$, where $A$ is the area of the region enclosing the quasiholes in question.
Working on an $L = 16 \ell_B$ cylinder and ensuring the quasiparticles remain at least a distance of $8 \ell_B$ apart, we find a statistical angle $\theta_{q = 3} = 2.0992$, compared to the prediction of $\frac{2 \pi}{3} \approx 2.0944$.
The computation takes about 1 minute.

	While the result is already well established for the Laughlin states,\cite{ArovasSchriefferWilczek1984} it would be worthwhile to explicitly calculate the non-abelian Berry connection for the Moore-Read quasiparticles.
As we have computed the form of the $Q$-matrices, we believe this would be tractable.

\section{Real-space entanglement spectrum}

	Finally, we present an algorithm for computing the real-space entanglement spectrum (RSES) of quantum Hall states on both finite and infinite cylinders.
In contrast to the orbital cut which divides the system into two sets of LLL orbitals,\cite{Haque2007}
	the real-space cut partitions the system into two regions of physical space.
Previously this was accomplished analytically for the free $\nu = 1$ case, \cite{Turner2010, Rodriguez:EntEntropyIQHE09}
	and numerically using Monte Carlo\cite{Rodriguez:RSES12} and large scale singular value decomposition (SVD) of explicit wave functions.\cite{Dubail2012a, Sterdyniak2012}
%
Our technique is not specific to the model wave functions, and provides a means for computing the RSES of non-model states calculated from DMRG.
As the scaling form of the entanglement spectrum only appears for the real-space cut, this may prove an important diagnostic for non-model states.
For simplicity, we assume a wave function in the LLL, and consider an entanglement cut running around the cylinder at $\tau = \tau_c$.
	
\begin{figure}[t]
	\subfigure[Split operation.]{ \label{fig:split} \includegraphics[width=0.4\textwidth]{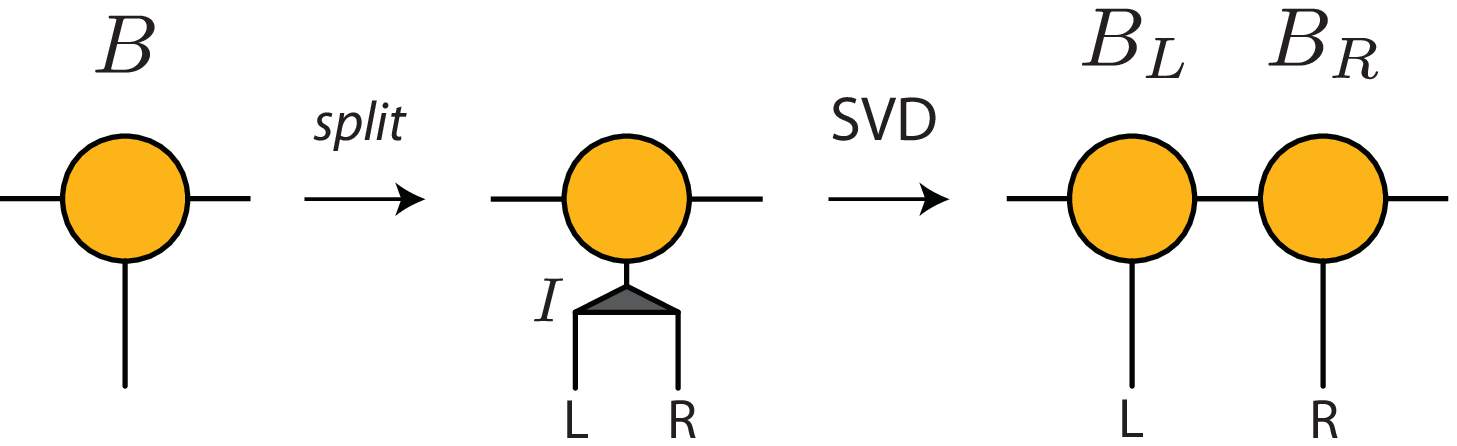} }
	\subfigure[Swap operation.]{ \label{fig:swap} \includegraphics[width=0.43\textwidth]{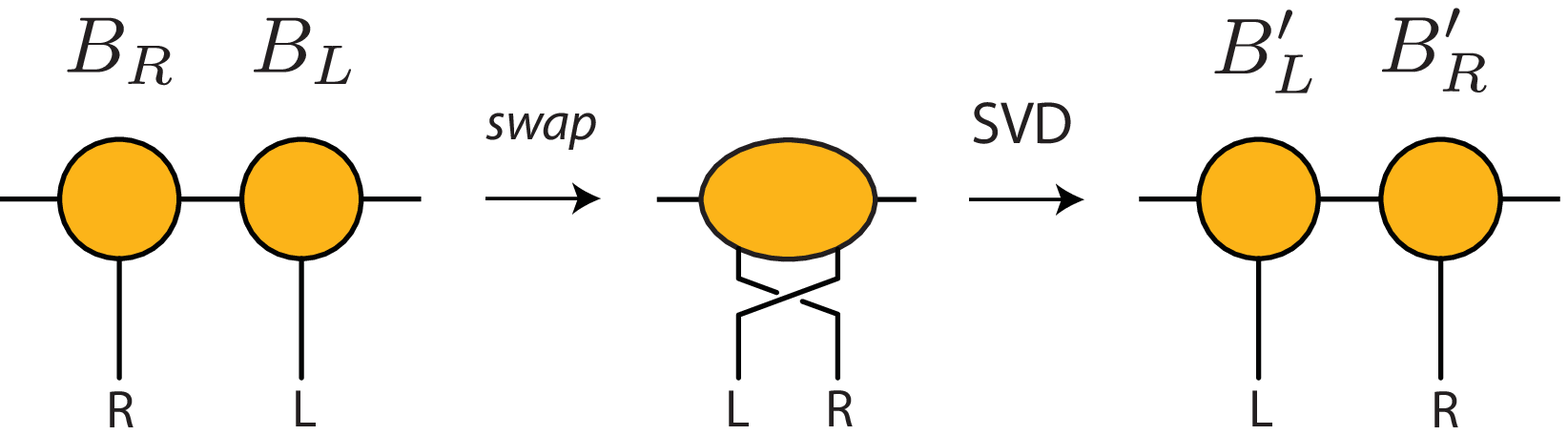} }
	\\[-3mm]
	\caption{%
		The splitting and swapping\cite{YYShiVidal-2006} procedure performed on an MPS.
		The two operations allow us to compute the real-space entanglement cut from an orbital MPS.
	}
	\label{fig:rs_split_swap}
\end{figure}

	The first step of the algorithm is to `split' each orbital $\varphi_n(z)$ [Eq.~\eqref{eq:LLL}] into components $\varphi_{n R/L}$ supported on the right and left of a cut at $\tau_c$,
\begin{align}
	\varphi_n(z) &= \theta(\tau - \tau_c) \varphi_n(z) + \theta(\tau_c - \tau) \varphi_n(z)	\notag\\
		&= g_{nL} \varphi_{nL}(z) + g_{nR} \varphi_{nR}(z).
\end{align}
We normalize the split orbitals by factors $g_{L/R}$ such that $\{\varphi_{nL}, \varphi_{nR} \}$ remains an orthonormal basis.
Expressing the state in terms of the `split' basis amount to appending isometries $I$ onto the $B$-matrices,
\begin{equation}
	B_{\alpha \beta}^m \to B^m_{\alpha \beta} I_m^{k l}, \quad
	I_m^{k l} = \sqrt{\binom{m}{k}} \, g_L^{k} \,  g_R^{l} \; \delta_{k+l, m}	,
\end{equation}
where the factors $g$ implicitly depend on the orbital location.
Because the orbitals $\varphi_n$ are exponentially localized about $\tau_n = \frac{2 \pi \ell_B^2}{L} n$, we can work at some fixed accuracy by splitting only the $M \sim \mathcal{O}\big(\frac{L}{\ell_B}\big)$ orbitals nearest to the cut.
In practice, we find $M = 1.5L/\ell_B$ is sufficient to obtain a converged spectrum.
As illustrated in Fig.~\ref{fig:split}, the affected $B$ matrices are then split using a SVD equivalent to the truncation step of time evolving block decimation (TEBD).\cite{PhysRevLett.98.070201}
As with TEBD, the splitting step preserves the `canonical' form of the MPS, implying that the bipartition about the new bond is a Schmidt decomposition.

	After the splitting step we have added $M$ $B$-matrices to the chain, with orbitals alternating between the left and right sides of the cut.
Choosing some particular bond to represent the location of the cut (usually the bond at the center of the set of sites we have split), we sort the MPS through a series of $\mathcal{O}(M^2)$ swapping procedures, bringing all indices associated to the left region to the left of the cut, and likewise for the right.
To accomplish this, we employ the swapping algorithm described in Ref.~\onlinecite{YYShiVidal-2006} to exchange each pair of neighboring sites in the MPS.
As illustrated in Fig.~\ref{fig:swap}, for each swap we form a two-site wave function, permute the right and left legs to bring them to the desired order, and then split the wave function using SVD to obtain a new pair of $B$-matrices.
Again, the canonical form of the MPS is preserved during this procedure, so after performing the required swaps the bond designated as the cut gives the real-space Schmidt decomposition.
	
	Depending on the initial bond dimension $\chi$, it may be necessary to truncate the new $B$-matrices by keeping only the largest singular values of the SVD.
It appears that the low-lying states are not affected by truncation of the highest lying states, but the convergence with increased $\chi$ should be checked on a case-by-case basis.

\section{Entanglement spectrum: orbital cut vs. real-space cut}
\label{sec:orb_vs_rs}
		
\begin{figure}[t]
	\includegraphics[width=80mm]{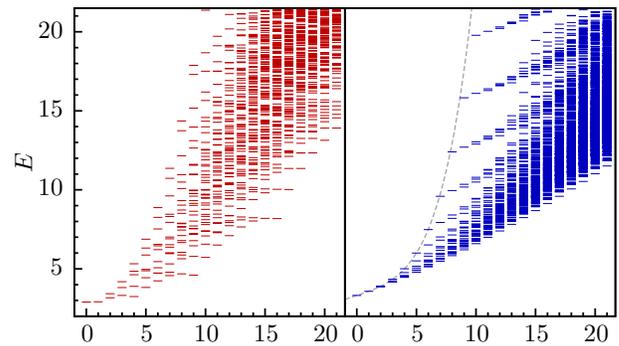}
	\caption{%
		Entanglement spectra (in the neutral charge sector) of the orbital (left) and real-space (right) cut of the $q=3$ Laughlin state at $L = 32\ell_B$.
		The energies $E$ are plotted against their momentum in units of $\Delta k = \frac{2\pi}{L}$.
		Both spectra have the counting 1, 1, 2, 3, 5, 7, 11, \textit{etc.}, consistent with that of a chiral boson CFT.
		However, the energy levels $E$ have vastly different quantitative behaviors in the two cases, which we investigate in Fig.~\ref{fig:q3_extrapolate_EE}.
		The dashed line on the right is of the form $v\epsilon(k) = vk [1 + u_2 k^2 + u_4 k^4]$, with $u_2, u_4$ fit from the highest level of each sector, which we associate with the state $a^\dagger_n\ket{0}$.
		The fits appear to rule out a similar term $u_1$, but larger sizes and a treatment of the `interactions' would be required to rule out $u_3$ if it is indeed absent. 
	}
	\label{fig:q3_ES}
\end{figure}

	In this section we study the scaling form of the RSES, then contrast it to the orbital spectrum.
As illustrated for the $q=3$ Laughlin state in Figs.~\ref{fig:q3_ES}, and for the MR state in Figs.~\ref{fig:q2+_ES}, the orbital and real-space cuts agree in their counting, which is that of the CFT, but differ in the scaling of the energy levels $E_i$ present in the spectrum.
	
	Kitaev and Preskill\cite{KitaevPreskill:TEE06} first noted that the known universal features of topological entanglement entropy could be explained if the \emph{energies} of the entanglement spectrum coincided with those of the chiral CFT. 
A physical argument was later provided in Ref.~\onlinecite{QiKatsuraLudwig2012}.
Recall that the states of the CFT are grouped into `families' associated with each primary field $\phi_h$,\cite{FendleyFisherNayak2007} in this context one per degenerate ground state on a cylinder,  and let $\hat{P}_{\phi_h}$ denote a projection operator onto the corresponding family.
The basic conclusion of Ref.~\onlinecite{QiKatsuraLudwig2012} was that the reduced \emph{real-space} density matrix of a topological state with gapless chiral edge modes takes the form
\begin{align}
	\hat{\rho}_L = \sum_{h} p_h \hat{P}_{\phi_h} e^{-v \hat{H} + \mathcal{O}(k \ell_B)}  \hat{P}_{\phi_h}.
\end{align}
Here $\hat{H}$ is the Hamiltonian of the CFT, which we will take to have velocity 1, so an `entanglement velocity' is included as a factor $v$. $\hat{H}$ is perturbed by more irrelevant boundary operators of order ${(k\ell_B)}^\delta$ for $\delta > 1$.
The coefficients $p_h$ depend on the degenerate ground state being considered. 
During the final preparation of this work, this scaling form was put on firm footing for the model FQH states.\cite{Dubail2012b}
	
	While the irrelevant operators generally introduce `interaction terms' to the entanglement Hamiltonian, to illustrate the expected behavior we consider the simplest type of correction, a dispersive term.
For the Laughlin state this takes the form
\begin{align}
	\tilde{E}_a &\equiv E_a - E_0  \sim v \left[ \sum_{n>0} \epsilon(k_n) a^\dagger_n a_{n} + \frac{2 \pi}{L} \frac{\hat{N}^2}{2 q}\right]	,\\
	\epsilon(k) &= k [1 + u_2 k^2 + u_4 k^4 + \cdots], \quad k_n = \frac{2 \pi}{L}n	.
\end{align}	
which accounts for the `branches' apparent in the real-space spectrum (Fig.~\ref{fig:q3_ES} right), each of which is associated with the presence of a new mode $a_n^\dag$.
The dispersion relation can be fit from the heights of these branches.
Note that only odd powers of $k$ can appear in the dispersion of a chiral boson.
In general, if the irrelevant perturbations descend from the identity boundary operator, only odd powers in $k$ should appear.

\begin{figure}[t]
	\subfigure[Real-space cut: plot of $\tilde{E}L/2\pi$ vs.\ $-1/L^2$.]{ \label{fig:rs3_EE} \includegraphics[width=80mm]{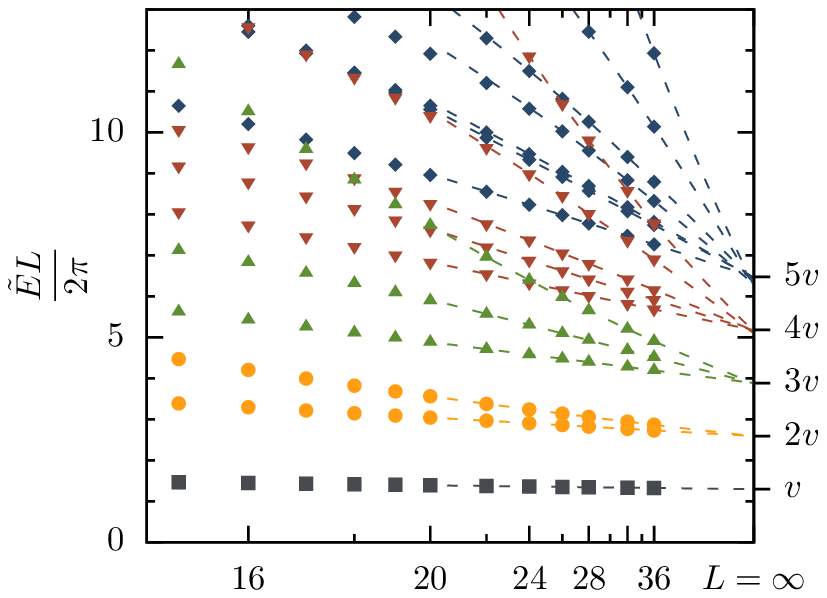} }
	\subfigure[Orbital cut: log-log plot of $\tilde{E}L/2\pi$ vs.\ $L$.]{ \label{fig:orb3_EE} \includegraphics[width=74mm]{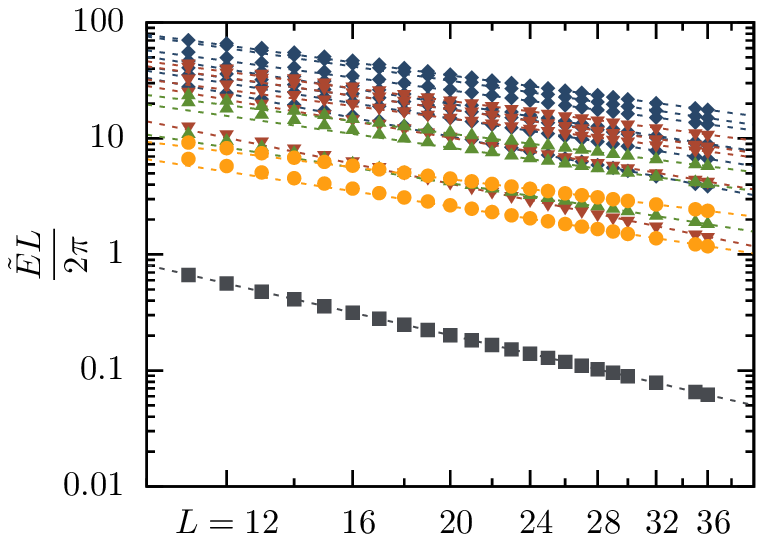} \quad\quad }
	\\[-3mm]
	\caption{%
		The relative entanglement energies for the real-space cut and orbital cut of the Laughlin state at $\nu = 1/3$.  (Data shown for the charge neutral sector, $L$ shown in units of $\ell_B$.)
		The states at different momenta are distinguished by their colored symbols.
		(a) For the real-space cut, we plot $\frac{\tilde{E} L}{2 \pi}$; the relative entanglement energy (relative to $E_0$) times the cylinder radius, as a function of $-1/L^2$.
		The energies are extrapolated to $L = \infty$ using a quadratic fit $t_0 + t_2 L^{-2} + t_4 L^{-4}$, and land on multiples of the entanglement velocity $v \approx 1.2956$.
		(b) For the orbital cut, we show $\frac{\tilde{E} L}{2 \pi}$ on a log-log plot, showing that the data has a linear behavior with negative slopes.
		The lines shown results from a linear fit to the last few data points.
		This demonstrates a power-law relation $\tilde{E}_a \propto L^{-\zeta_a}$ with $\zeta_a > 1$.
	}
	\label{fig:q3_extrapolate_EE}
\end{figure}
In order to accurately extract the entanglement velocity, we consider the scaling of the shifted spectrum $\frac{\tilde{E} L}{2 \pi}$ with increased $L$. Based on these scaling ideas, a state with momentum $k = \frac{2 \pi}{L}(n_\phi + n_\chi)$ should have an energy
\begin{multline}
	\label{eq:fit_form}
	\frac{\tilde{E}_a L}{2 \pi} = v_\phi ( \Delta_{\phi h} +  n_\phi )  + v_\chi ( \Delta_{\chi h}  +  n_\chi ) \\
			\quad + t_{2a} L^{-2} + t_{4a} L^{-4} + \cdots
\end{multline}
where $n_\phi$ and $n_\chi$ are integers corresponding to the momenta of the $\mathrm{U(1)}$ and Majorana sectors.
The offsets $\Delta_{\phi/\chi h}$ are the scaling dimensions of the highest weight state in the sector, which depends on the bond and number sector in question (for the Laughlin states, it is $\frac{N^2}{2 q}$).
For the MR case we have included a detailed exposition of this structure in Appendix~\ref{app:MR_counting}.

Focusing on the identity sector $\Delta_h = 0$ of the \emph{real-space} $q=3$ Laughlin cut, Fig.~\ref{fig:rs3_EE} tracks the scaled relative entanglement energy levels $\frac{\tilde{E} L}{2 \pi}$ as a function of $L^{-2}$, extrapolating their value as the circumference approaches infinity. 
As indicated by the right-most tics of the figure, $\frac{\tilde{E} L}{2 \pi}$ approaches $n_\phi v_\phi$ for large $L$, where $v_\phi \approx 1.2956$. The data clearly confirms that the real-space entanglement spectrum approaches a linear dispersion with fixed velocity, and the success of the fit justifies the absence of $L^{-1}$ and $L^{-3}$ perturbative terms.
We have tabulated the velocities for the $q = 1$, $3$, $5$ and $7$ in Tab.~\ref{tab:vel_gamma}.
The relation $v_{q=1} / v_{q=3} \approx \sqrt{3}$ noted previously\cite{Sterdyniak2012} appears not to continue to higher $q$.

\begin{figure}
	\includegraphics[width=85mm]{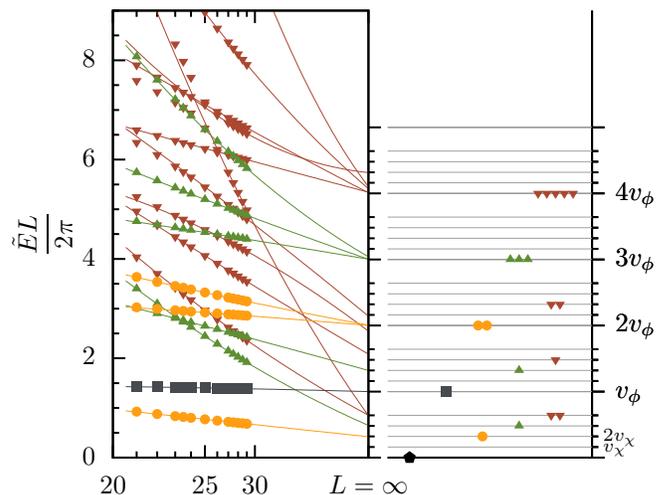}
	\caption{%
		Extrapolating the entanglement energies $\tilde{E} L / 2\pi$ for the Moore-Read state at $\nu = 1/2$, using a cut associated with counting 1, 1, 3, 5, 10, \textit{etc.} for $N = 0$ charge sector (\textit{cf.}\ App.~\ref{app:MR_counting}).
		(Left)
		Here we show that the energies for the first three momenta extrapolate to integral combinations of $v_\phi$ and $v_\chi$, the velocities of the chiral boson and Majorana mode respectively.
		The major tics on the vertical axis labels multiples of $v_\phi$, the minor tics label combinations $n_\phi v_\phi + n_\chi v_\chi$ for integers $n_\phi$ and $n_\chi$ (and $\ell_B$ set to 1).
		States with momentum $4 \frac{2 \pi}{L}$ extrapolate near, but not exactly, to the theoretical prediction, which we attribute to smallness of the system sizes.
		The superimposed lines are quadratic fits over the largest few circumferences, extrapolating to give $v_\phi \approx 1.33$, $v_\chi \approx 0.21$.
		(Right)
		The theoretical placement of the energy levels for the state.
		Here the boson counting (1,1,2,3,5) and the Majorana counting (1,0,1,1,2) are apparent.
		See App.~\ref{app:MR_counting} for a detailed explanation of the counting in this plot and data for other charge sectors.
	}
	\label{fig:mr_extrapolate_EE}
\end{figure}

This same technique can be used for more complicated wave functions such as the Moore-Read state, as shown in Fig.~\ref{fig:mr_extrapolate_EE}.
We extrapolate the velocities of both the charge and neutral modes to be $v_\phi \approx 1.33$ and $v_\chi \approx 0.21$ respectively.
We note that the extrapolation is only possible for sufficiently large circumferences $L \gtrsim 20\ell_B$, which is well within reach using the MPS representation of the wave function.

Figure~\ref{fig:orb3_EE} shows that in the orbital-cut, $\tilde{E}$ does not extrapolate to the CFT linear dispersion.
Rather, they appear to follow power law decays $\tilde{E}_a \sim L^{-\zeta_a}$ with different $\zeta_a$ for each state $a$.
For example, the fit for $k = \frac{2\pi}{L}$ gives $\zeta \approx 3.0$, while $\zeta \approx 2.3$, $2.1$ for the two set of states at $k = 2\frac{2\pi}{L}$.
(In the real-space case, $\zeta = 1$ for all the levels.) 
Unfortunately, the range of data available is insufficient to draw any conclusions.

Finally, we note that one can extract the topological entanglement entropy in either types of cut.
This was shown in Fig.~\ref{fig:gamma_fit} where we used both the entropy $S$ and the zero momentum state $E_0$ as a function of $L$.
For each $L$ we perform a windowed fit; presenting the intercept of the best line fit through the neighboring points.
While it is possible to extract $\gamma$ from any of the four computed quantities, we can see that the real-space cut is less oscillatory than the orbital cut.
At the same time, using the orbital cut $E_0$ seems to give a much better convergence of $\gamma$ than any of the other methods,
\textit{i.e.}, the system size $L$ required to computed $\mathcal{D} = e^{\gamma}$ via the orbital $E_0$ to accuracy $\pm0.5$ is the smallest.
Table~\ref{tab:vel_gamma} lists the entanglement velocities and TEE for various Laughlin states and the $q=2$ MR state,
	as well as the velocities extracted via the method used in Figs.~\ref{fig:rs3_EE} and \ref{fig:mr_extrapolate_EE}.

\section{Conclusion}

We have shown how the CFT structure in model FQH wave functions enable us to represent them as matrix product states.
These MPSs can be evaluated numerically on an infinite cylinder; the distinct advantages of this geometry, as well as the efficiency of the MPS, allow us to study in detail the scaling properties of the Moore-Read entanglement spectrum, including a definitive identification of the $\mathrm{U(1)}$ and Majorana modes and their velocities.

There are several future directions.
The MPS representation is well suited for studying the screening properties of the states as well as their Berry connections, so it would be valuable to numerically implement the MR quasiholes in order to verify various screening arguments.\cite{Read2009, BondersonNayak2011}
As we have noted, our construction also generalizes to other topological phases whose model states can be expressed as a correlation function of a lower dimensional field theory.
The resulting picture is strikingly similar to the `entanglement renormalization' classification of 1D phases exemplified in the AKLT state.
In particular, it would appear that the fixed points of the entanglement renormalization scheme may be interpreted as some form of fixed point for the auxiliary field theory when expressed as a tensor network -- for topological phases, a massless fixed point, while for trivial phases, a massive fixed point.
Making this connection precise would be an intriguing development.

We would like to acknowledge helpful conversations with Joel~E. Moore, Tarun Grover, Frank Pollmann, Sid Parameswaran, and J\'er\^ome Dubail,
	as well as support from NSF GRFP Grant No.~DGE 1106400 (MZ) and NSF DMR-0804413 (RM).

\clearpage
\appendix

\section{\texorpdfstring
	{Evaluation of $B$-matrices for Laughlin states}
	{Evaluation of B-matrices for Laughlin states}}
\label{app:eval_B}

	Here we provide more detail on the precise form of the Laughlin MPS and its numerical implementation.
The mode expansion of the chiral boson is
\begin{align}
	\phi(w) &= \sum_{n\neq0} \frac{w^{-n}}{\sqrt{|n|}} a_n + \phi_0 + \frac{\pi_0}{i} \log(w), \\
	& [\phi_0, \pi_0] = i, \quad [a_n, a_m] = \delta_{n+m}.	\notag
\end{align}
The field is composed of the fluctuating part $\phi'(w)$ and the `zero mode',
\begin{align}
	\phi(w) = \phi'(w) + \phi_0 + \frac{\pi_0}{i} \log(w)	\,.
\end{align}
The states of the fluctuating sector can be labeled by occupation numbers, which we denote by a string of positive integers $P$.
For example, $\ket{0}$ denotes the ground state, $\ket{221} = \frac{1}{\sqrt{2}} a_2^\dag a_2^\dag a_1^\dag \ket{0}$, \textit{etc}.
We define $|P|$ to be the total momentum of the fluctuations in $\ket{P}$, given by the sum of the integers.
The states of the zero-mode sector are labeled by the eigenvalues of $\pi_0$.
For convenience, we define `charge' by $ \hat{N} = \sqrt{q} \pi_0$, chosen such that the electron has charge $q$.
The states of the zero-mode sector are labeled by $\ket{N}$, so the full CFT is then spanned by $\ket{P, N}$. 

	Treating first the `free' evolution $U$, we find
\begin{align}
	H &= \frac{2 \pi}{L} \left[ |P| + \frac{1}{2 q} N^2 \right]	\,,\\
	U(\delta\tau)_{P,N; P', N'} &= \delta_{P, P'} \delta_{N, N'}
			e^{- \left( \frac{2 \pi \ell_B}{L}\right)^2 \left[ |P| + \frac{1}{2 q} N^2 \right]}.
\end{align}
Now we calculate the on-site term $T$, first by converting from the coherent state form $T[\psi]$ to the occupation basis, $T^m$:
\begin{align}
	T[\psi] &= e^{- \frac{i}{2 \sqrt{q}} \phi_0}  e^{ \hat{\mathcal{V}}_{0} \psi} e^{- \frac{i}{2 \sqrt{q}} \phi_0}	\notag \\
	&= \sum_m  T^m {(m!)}^{3/2}  \psi^m. \\
	T^m &\equiv \frac{1}{\sqrt{m!}}  e^{- \frac{i}{2 \sqrt{q}} \phi_0}  \big(\hat{\mathcal{V}}_0\big)^m e^{-\frac{i}{2\sqrt{q}} \phi_0}	.
\end{align}
We next compute the matrix elements of the vertex operator,
\begin{align}
	&	\bra{P, N}\hat{\mathcal{V}}_{0}\ket{P', N'} 	\notag\\	
		&\quad	= \bra{P, N}{e^{i \sqrt{q} \phi'(w) + i \sqrt{q} \phi_0 + N \log(w)] }}\ket{P', N'}	.
\end{align}
The zero-mode part depends only on $N$,
\begin{align}
	\bra{N}{e^{i \sqrt{q} \phi_0 + \hat{N} \log(w) } }\ket{N'} = \delta_{N - N',  q}\,  w^{{N+N'}/2}.
\end{align}
The fluctuating part depends only on the oscillators $\ket{P}$, so we define
\begin{align}
	A^n_{P, P'} = \bra{P}{\oint\! \frac{dw}{2 \pi i} \,  w^{-n-1}  e^{i \sqrt{q} \phi'(w)}}\ket{P'}	.
\end{align}
Hence $A^n$ is simply the $n$th coefficient of a Taylor expansion in $w$.
The matrices $A$ are non-zero only for $P - P' = -(N + N')/2$, due to momentum conservation.
Numerically, we impose a cutoff $\Lambda$ such that we only keep states $\ket{P}$ with $|P| \leq \Lambda$, which allows us to evaluate $A$ for only a finite number of states.
The time to compute $A$ is proportional to its number of entries, so the construction of the MPS is an insignificant part of the computational cost (\textit{i.e.}, compared to matrix multiplication).
Combining the zero-mode and fluctuations,
\begin{align}
	\bra{P, N}{\hat{\mathcal{V}}_{0}}\ket{P', N'} = A^{-\frac{N + N'}{2}}_{P P'}  \delta_{N - N', q}. 
\end{align}
Finally, the sandwiching background charge contributes $e^{-\frac{i}{\sqrt{q}} \phi_0} = \delta_{P, P'} \delta_{N - N', -1}$ to each site.

	Focusing on the case of fermions where there is at most one particle per orbital,
\begin{subequations}\label{eq:Tmatrices}\begin{align}
	T^{0}_{P,N; P',N'} &= \delta_{P, P'} \delta_{N - N', -1}		&\textrm{(unoccupied),}&	\\
	T^{1}_{P,N; P',N'} &= A^{-\frac{N + N'}{2}}_{P P'}  \delta_{N - N',  q - 1}	&\textrm{(occupied).}&
\end{align}\end{subequations}
For case of bosons, the higher occupation states involve products of the $A$'s.

	The $q$ fold ground state degeneracy of the Laughlin states can be seen by noting that on a particular bond, $e^{2 \pi i N / q}$ is a constant, and can be chosen to take one of $q$ values.

\section{\texorpdfstring
	{Evaluation of $B$-matrices for Moore-Read state}
	{Evaluation of B-matrices for Moore-Read state}}
\label{app:eval_B_MR}

	The CFT associated with the Moore-Read state is a tensor product of a chiral boson $\phi$ and a Majorana mode $\chi$.
We first give a brief review of the structure of the chiral Majorana CFT on a cylinder.\cite{Ginsparg}
The states form four sectors according to their boundary condition (bc), (periodic `\textbf{P}' or antiperiodic `\textbf{AP}') and number parity (even `$+1$' or odd `$-1$').
We denote the lowest energy states of these four sectors by
	`$\ket{\bbone}$' for \textbf{AP}/1,
	`$\ket{\chi}$' for \textbf{AP}/$-1$,
	`$\ket{\sigma}$' for \textbf{P}/1
	and `$\ket{\mu}$' for \textbf{P}/$-1$.
In the periodic sector the Majorana has modes $\chi_{n} : n \in \mathbb{Z}$, while in the anti-periodic sector it has modes $\chi_{m} : m \in \mathbb{Z} + \tfrac{1}{2}$.
The states of the \textbf{P}/\textbf{AP} sectors can be obtained by acting with the \textbf{P}/\textbf{AP} modes $\chi_{-m}$ on  $\ket{\sigma}$/$\ket{\bbone}$ respectively.
Within a given sector, the states can then be labeled by a string of numbers $P_\chi$; they are either integers or half-integers depending on the bc, and do not repeat because of the fermionic statistics.
Letting $|P_\chi|$ denote the total momentum of the Majorana, 
\begin{equation}
	H_\chi = \frac{2 \pi}{L} \left[ |P_\chi| + \Delta  \right]
\end{equation}
where $\Delta = \{0, \frac{1}{16} \}$ for the \textbf{AP} and \textbf{P} sectors respectively, though $\Delta$ can be ignored as it only changes the normalization of the state.
	
	The operator $e^{i \sqrt{q} \phi(z)} \chi(z)$ must be periodic in $z$ at the location of the Landau orbitals $\tau_n$ (our choice of gauge has a twist boundary condition in between).
This introduces a constraint between the zero mode of the boson, $\hat{N} = \sqrt{q} \hat{\pi}_0$, and the boundary condition of the Majorana.
We find that for $q$ even (the fermionic case), at the \emph{bond} of the MPS the CFT boundary condition is such that if the Majorana is in  \textbf{P}, we must have $N \in \mathbb{Z} + \frac{1}{2}$, while for \textbf{AP}, we must have $N \in \mathbb{Z}$.
The boundary conditions will correspond to different degenerate ground states, with four states of type \textbf{AP} and two of type \textbf{P}, for a total of six on the infinite cylinder (for a torus, the \textbf{P}  sector acquires an additional two states depending on the parity of the electron number).\cite{ReadGreen:p+ipFQHE00}
	
	The total energy of the combined CFT is
\begin{equation}
	H = \frac{2 \pi}{L} \left[ |P_\chi| + |P_\phi| + \frac{1}{2 q} N^2 \right]
\end{equation}	
with $|P_\phi|$ and $N$ arising for the boson.
As for the Laughlin state, $U = e^{-\delta \tau H}$ is diagonal if we work in the occupation basis.
Constructing the $T$ matrices proceeds as for the Laughlin case, but we must include the Majorana sector in the computation of $\hat{\mathcal{V}}_0 = \oint\!\frac{dw}{2 \pi i} w^{-1} \chi(w) e^{i \sqrt{q} \phi(w) }$.
Letting
\begin{equation}
	\chi^m_{P_\chi, P'_\chi} = \bra{P_\chi} \chi_{m} \ket{P'_\chi}
\end{equation}
denote the matrix elements of the Majorana operators, the required matrix element is
\begin{multline}
	\bra{P_\phi, P_\chi, N}\hat{\mathcal{V}}_{0}\ket{P'_\phi, P'_\chi, N'}	\\
		=  \sum_m \chi^{-m}_{P_\chi, P'_\chi} A^{m-\frac{N + N'}{2}}_{P_\phi, P'_\phi} \, \delta_{N - N',  q}	\,,
\end{multline}
with $A$ defined as for the Laughlin case. 

	For fermions, where there is at most one particle per orbital,
\begin{subequations}\label{eq:Tmatrices_MR}\begin{align}
	&	T^{0}_{(P_\phi,P_\chi,N), (P',P_\chi',N')}		\\\notag
		&\quad= \delta_{P_\phi, P_\phi'} \delta_{P_\chi, P_\chi'} \delta_{N - N', -1}	&&\textrm{(unoccupied),}	\\
	&	T^{1}_{(P_\phi,P_\chi,N), (P',P_\chi',N')} 	\\\notag
		&\quad= \sum_m \chi^{-m}_{P_\chi, P'_\chi} A^{m-\frac{N + N'}{2}}_{P_\phi, P'_\phi}  \delta_{N - N',  q - 1}	&&\textrm{(occupied).}
\end{align}\end{subequations}
For the case of bosons, the higher occupation states involve products of the $\mathcal{V}_0$'s.

\begin{figure}[t]
	\subfigure[]{ \label{fig:AProot} \includegraphics[width=0.25\textwidth]{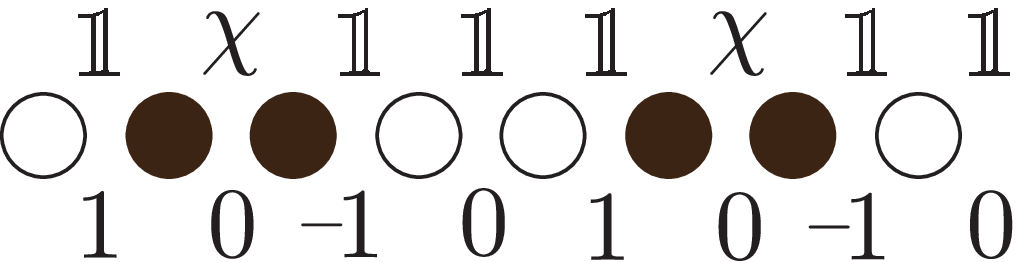} }
	\subfigure[]{ \label{fig:Proot} \includegraphics[width=0.25\textwidth]{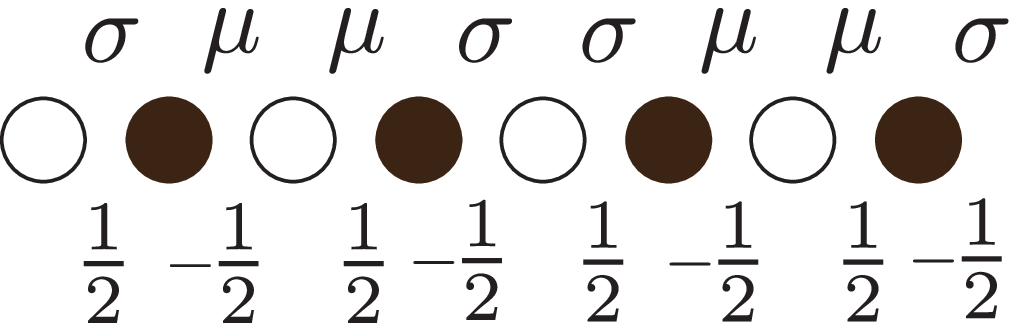} }
	\caption{%
		The thin torus orbital wave functions for the 
		(a) antiperiodic and (b) periodic sectors.
		Each site corresponds to an orbital, which is either filled (black) or empty (white).
		Each bond has only a single state of the CFT, which we decompose into the Majorana part, shown above the bond, and zero mode of the boson $N$, shown below the bond.
	}
	\label{fig:MRroot}
\end{figure}

	The $3 q$ fold ground state degeneracy of the MR states can be seen by first choosing a bc sector for the Majorana, \textbf{P} or \textbf{AP}.
In the \textbf{AP} sector, on any given bond $(-1)^F e^{i \pi N / q}$ is constant, with $N \in \mathbb{Z}$ and $F$ the Majorana number. The quantity has $2 q$ allowed values, each leading to a distinct state.
$(-1)^F e^{i \pi N / q}$ is also constant in the \textbf{P} sector, where $N \in \mathbb{Z} + \frac{1}{2}$.
However, here the $2 q$ values only lead to $q$ distinct states.
This is because while inserting the Majorana zero-mode $\chi_0$ at past infinity changes the assignment of $F$, it does \emph{not} actually change the physical state.
As illustrated in Fig.~\ref{fig:MRroot}, this can be understood as a simple relabeling $\mu \leftrightarrow \sigma$, which is equivalent and so produces the same state.

In the small $L$ limit, these $3q$ states evolve into `thin torus' wave functions.\cite{Bergholtz2006}
In this limit, we restrict the Majorana CFT to the states $\{ \ket{\bbone}, \ket{\chi}, \ket{\sigma}, \ket{\mu} \}$, and fix $|P_\phi| = 0$ for the boson, which projects onto the charges $N = \{ -1, -\frac{1}{2}, 0, \frac{1}{2}, 1 \}$.
The six resulting states are precisely the `highest weight' states of the CFT, as illustrated in Fig.~\ref{fig:MRroot}.

\section{The counting of the Moore-Read state}
\label{app:MR_counting}

As explained in the last section, the chiral Majorana CFT may be separated into four sectors, by periodicity of the boundary as well as the particle number parity.

\begin{subequations}
In the periodic sectors $\sigma$ and $\mu$, the excitations have momenta which are integral multiples of $\Delta k = \frac{2\pi}{L}$,
hence the counting of level $n$ is the number of partitions of $n$ into an even/odd number of distinct non-negative integers.
The number of states at momenta $0, \Delta k, 2\Delta k, ...$ are as follows,%
	\footnote{The sequence \eqref{eq:counting_maj_p} is given at \href{http://oeis.org/A000009}{oeis.org/A000009}.}
\begin{align}
	&	\mu, \sigma:		& 1, 1, 1, 2, 2, 3, 4, 5, 6, 8, 10, 12, 15, 18, ...	\label{eq:counting_maj_p}
\end{align}
(Because of the presence of the zero-momentum mode, the countings of the two \textbf{P} sectors are identical.)

In the antiperiodic sectors the excitations have momenta which are integer-plus-half multiples of $\Delta k$,
	or in other words, twice the momentum is always an odd multiple of $\Delta k$.
Hence in the $\bbone$ sector the counting of level $n$ is given by the partitions of $2n$ into positive odd integers.
	\footnote{The fact that we require an even number of integers in the partition is automatically enforced.
		The sequence \eqref{eq:counting_maj_1} is given at \href{http://oeis.org/A069910}{oeis.org/A069910}.}
\begin{align}
	&	\bbone:		& 1, 0, 1, 1, 2, 2, 3, 3, 5, 5, 7, 8, 11, 12, ...	\label{eq:counting_maj_1}
\end{align}
The same definition also hold for the $\chi$ sector, with counting as follows,%
	\footnote{The sequence \eqref{eq:counting_maj_chi} is given at \href{http://oeis.org/A069911}{oeis.org/A069911}.}
\begin{align}
	&	\chi:		& 1, 1, 1, 1, 2, 2, 3, 4, 5, 6, 8, 9, 12, ...	\label{eq:counting_maj_chi}
\end{align}
Note that since there are an odd number of excitations, the lowest energy state is $\ket{\chi} = \chi_{1/2} \ket{\bbone}$ with momentum $\frac{1}{2}\Delta k$.
The counting in this sector corresponds to the number of states at momenta $\frac{1}{2}\Delta k, \frac{3}{2}\Delta k, \frac{5}{2}\Delta k, ...$.
\end{subequations}

\begin{figure}[t]
	\includegraphics[width=80mm]{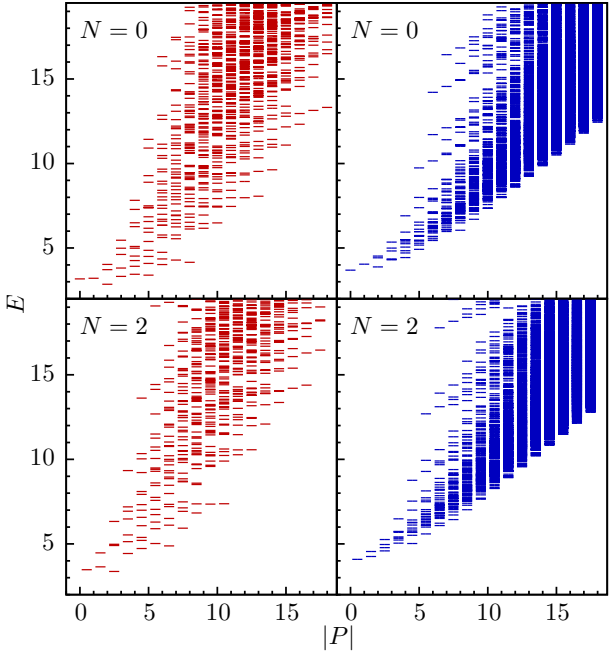}
	\caption{%
		The entanglement spectra of the $q=2$ Moore-Read state at $L = 25\ell_B$,
			with the orbital (left) and real-space (right) cut,
			in the $N = 0$ (top) and $N = 2$ (bottom) charge sectors.
		In the orbital case, the cut takes place on the bond with $\ket{\bbone}$ Majorana and $N = 0$ boson state
			in the thin torus limit (\textit{cf.}\ Fig.~\ref{fig:MRroot}).
		The real-space cut takes place at the $\tau$ centered on that bond.
		For $N = 0$, the counting of the states is 1, 1, 3, 5, 10, 16, \textit{etc},
			while for $N = 2$, the counting is 1, 2, 4, 7, 13, \textit{etc}.
	}
	\label{fig:q2+_ES}
\end{figure}

Combined with the chiral boson, the counting of the Moore-Read edge spectra are%
\cite{WenMooreReadEdge93}\textsuperscript{,}
\footnote{The sequence \eqref{eq:counting_mr_p} is given at \href{http://oeis.org/A015128}{oeis.org/A015128}.
	The sequences \eqref{eq:counting_mr_1} and \eqref{eq:counting_mr_chi} interlaced together is given at \href{http://oeis.org/A006950}{oeis.org/A006950}.
	}
\begin{subequations}\begin{align}
	\bbone	&:		& 1, 1, 3, 5, 10, 16, 28, 43, 70, ...	\label{eq:counting_mr_1}\,,	\\
	\chi	&:		& 1, 2, 4, 7, 13, 21, 35, 55, 86, ...	\label{eq:counting_mr_chi}\,,	\\
	\mu, \sigma	&:		& 1, 2, 4, 8, 14, 24, 40, 64, 100, ...	\label{eq:counting_mr_p}\,.
\end{align}\end{subequations}
(Again, in the $\chi$ sector, the momenta are shifted by $\frac{1}{2}\Delta k$.)
Figure~\ref{fig:q2+_ES} shows the orbital and real-space cut of the $q=2$ MR state giving the $\bbone$ sector.

\begin{figure}[t]
	\includegraphics[width=85mm]{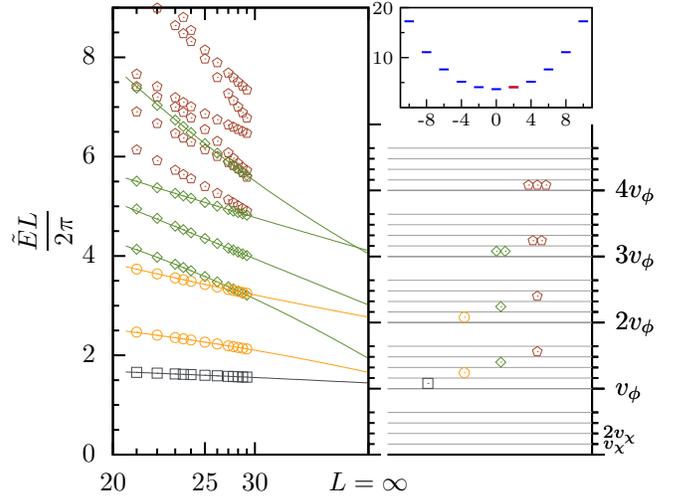}
	\caption{%
		Extrapolating the entanglement energies $\tilde{E} L / 2\pi$ for the Moore-Read state at $\nu = 1/2$.
		The real-space cut is physically centered on the bond with $\bbone$ sector (in the thin torus limit),
			but the data is shown in the $N = 2$ charge sector, and hence the counting matches that of the $\chi$ sector.
		\\[1.5mm]
		(Left) The energy states at momenta $\frac{\pi}{L}$, $3\frac{\pi}{L}$, $5\frac{\pi}{L}$, $7\frac{\pi}{L}$ are labeled by their shape, with their values extrapolated to $L = \infty$.
		(Right) The theoretical energy levels of the MR state for the $\chi$ sector.
		(Upper right inset) The least-momentum state in each charge sector $N$ at $L = 25\ell_B$, with the sector of interest marked red.
		\\[1.5mm]
		The lowest states extrapolates to $v_\phi + \frac{1}{2}v_\chi$, consistent with Eq.~\eqref{eq:fit_form} (using $\Delta_{\phi h} = \frac{N^2}{2q}=1$).
		(States with momenta $7\frac{\pi}{L}$ do not extrapolate to their theoretical values due to insufficient system sizes.)
		Contrast this plot to the $N=0$ sector of Fig.~\ref{fig:mr_extrapolate_EE}.
	}
	\label{fig:mr_extrapolate_EE_Q1}
\end{figure}

Notice that in the \textbf{AP} case, the Majorana sector alternates between $\bbone$ and $\chi$ sectors whenever an orbital is filled (see Fig.~\ref{fig:MRroot}).
Hence the entanglement spectrum with different charges would also alternate between the countings \eqref{eq:counting_mr_1} and \eqref{eq:counting_mr_chi}, shown clearly in Fig.~\ref{fig:q2+_ES}.
Figure~\ref{fig:mr_extrapolate_EE_Q1} shows the relative entanglement energies of the $q=2$ MR state at the $N=\pm q$ charge sectors; contrast this to the $N=0$ sector of Fig.~\ref{fig:mr_extrapolate_EE}.

\section{\texorpdfstring
	{Evaluation of $\mathcal{Q}$-matrices for Laughlin and Moore-Read quasiholes}
	{Evaluation of Q-matrices for Laughlin and Moore-Read quasiholes}}
\label{app:eval_Q_qh}
	Evaluation of the $Q$-matrices for the Laughlin state can be done in a similar manner to the bulk $B$-matrices, but  omitting the contour integration:	
\begin{align}
	Q_{P, N; P', N'} &= \delta_{N - N',  1}\,   (s w)^{\frac{N+N'}{2q}} \bra{P} e^{i  \phi'(w) / \sqrt{q}}\ket{P'}
\end{align}
where $w = e^{-\frac{2 \pi i}{L} \eta_x}$ and $s = \pm 1 $ for bosons or fermions respectively. Note that the momentum is no longer conserved.

	The Moore-Read case is more complex.
In the context of the Majorana CFT, the Ising order and disorder fields $\sigma, \mu$ are `twist' fields, interpolating between \textbf{AP} and \textbf{P} periodic bc's.
We take the point of view that the fields $\sigma$ and $\mu$ have fixed fermion parity $+1$ and $-1$ respectively.
The resulting fusion rules are
\begin{subequations}\begin{align}
	[\sigma][\sigma] = [\mu][\mu] &= [\chi][\chi] = [\bbone], \\
	[\mu][\sigma] = \chi, \, \,  [\mu][\chi] &= [\sigma], \, \, [\sigma][\chi] = [\mu].
\end{align}\end{subequations}
In this approach, there are \emph{two} possible quasihole insertions, $\sigma(\eta) e^{i \phi/\sqrt{q}}$, and $\mu(\eta) e^{i \phi/\sqrt{q}}$. As $\chi_0 \sigma \sim \mu$, this is a direct realization of the picture in which each vortex has a Majorana zero-mode. The non-trivial vector space of quasihole excitations arises from the freedom of choosing $\sigma$ or $\mu$, subject to the constraint that they fuse properly to the vacuum.

  Fortunately the techniques for evaluating  matrix elements of the type
\begin{align}
	\sigma_{P_\chi, P'_\chi} = \bra{P_\chi}{\sigma(0)}\ket{P'_\chi}
\end{align}
have already been developed in the `truncated-fermionic-space-approach' to the perturbed Ising CFT.\cite{YurovZamolodchikov1991} 

Consider, for example, the \textbf{AP} to \textbf{P} case.
Arbitrary states can be built by acting with the modes $\chi_{-n}$, so without loss of generality we consider the matrix element
\begin{align}
	\bra{\sigma}\prod_{\{m_i \in P_\chi \}} \chi_{m_i} \sigma(\eta) \prod_{\{n_i \in P'_\chi \}} \chi_{-n_i} \ket{\bbone}.
\end{align} 
The chief technical result of Ref.~\onlinecite{YurovZamolodchikov1991} Eqs.~{2.9-2.13} is that there exists an easily computed matrix $C(\eta)$ such that
\begin{align}
	\bra{\sigma} \cdots \chi_m \sigma(\eta) \cdots \ket{\bbone} = \bra{\sigma} \cdots  \sigma(\eta) C_{mn}(\eta)\chi_{n} \cdots \ket{\bbone}.
\end{align}
After commuting all $\chi$ across the insertion, the Majoranas are brought to normal ordered form, reducing the problem to Wick contractions and the matrix elements
\begin{subequations}\begin{align}
	\bra{\sigma}\sigma \ket{\bbone} &= C_{\sigma \sigma 1}  \\
	\bra{\sigma}\mu \ket{\chi} &= C_{\sigma \mu \chi} \quad \textit{etc.}
\end{align}\end{subequations}
We have not as of yet implemented the MR quasiholes numerically, which would be a worthwhile check given the subtleties of this case.

%

\hbadness=10000
\bibliography{qhmps}
\end{document}